\documentclass[usenatbib]{mnras}
\usepackage{natbib}
\usepackage{newtxtext,newtxmath}
\usepackage{iftex}
\ifxetex
  \usepackage[silent]{xeCJK}
\else
  \usepackage{CJKutf8}
\fi
\usepackage[T1]{fontenc}

\DeclareRobustCommand{\VAN}[3]{#2}
\let\VANthebibliography\thebibliography
\def\thebibliography{\DeclareRobustCommand{\VAN}[3]{##3}\VANthebibliography}

\usepackage{graphicx}	
\usepackage{amsmath}	
\usepackage{amssymb}	
\usepackage{soul}
\usepackage{listings}   
\usepackage[dvipsnames]{xcolor}     
\usepackage{xspace}     
\usepackage{orcidlink}

\usepackage{mathtools}
\newcounter{mysubequations}

\renewcommand{\themysubequations}{(\roman{mysubequations})}

\newcommand{\mysubnumber}{\refstepcounter{mysubequations}\themysubequations}

\title[Stellar parameters from Gaia XP spectra]{Parameters of 220 million stars from Gaia BP/RP spectra}

\ifxetex 
\else
\begin{CJK*}{UTF8}{gbsn}
\fi
\author[Zhang, Green \& Rix]{
    Xiangyu Zhang (张翔宇) \orcidlink{0000-0003-3112-3305}$^{1}$\thanks{E-mail: xzhang@mpia.de},
    Gregory M. Green \orcidlink{0000-0001-5417-2260}$^{1}$,
    Hans-Walter Rix \orcidlink{0000-0003-4996-9069}$^{1}$
    \\
    $^{1}$Max Planck Institute for Astronomy,
    K\"onigstuhl 17, D-69117 Heidelberg, Germany
}
\ifxetex
\else
\end{CJK*}
\fi

\date{Accepted XXX. Received YYY; in original form ZZZ}

\pubyear{TBD}

\begin{document}
    \label{firstpage}
    \pagerange{\pageref{firstpage}--\pageref{lastpage}}

    \ifxetex
      \maketitle
    \else
      \begin{CJK*}{UTF8}{gbsn}
      \maketitle
      \end{CJK*}
    \fi

    \newcommand{\feh}{\mathrm{[Fe/H]}}
\newcommand{\teff}{T_{\rm eff}}
\newcommand{\logg}{\log{g}}
\newcommand{\type}{\vec{\Theta}}
\newcommand{\dm}{$\mu$}
\newcommand{\ebv}{\ensuremath{\mathrm{E}\left( B \! - \! V \right)}\xspace}

\newcommand{\Gaia}{\textit{Gaia}\xspace}

\newcommand{\change}[1]{#1} 
    
    
    \tolerance=1
    \emergencystretch=\maxdimen
    \hyphenpenalty=10000
    \hbadness=10000
    
    \begin{abstract}
We develop, validate and apply a forward model to estimate stellar atmospheric parameters ($\teff$, $\logg$ and $\feh$), revised distances and extinctions for 220 million stars with XP spectra from \Gaia DR3. Instead of using \textit{ab initio} stellar models, we develop a data-driven model of \Gaia XP spectra as a function of the stellar parameters, with a few straightforward built-in physical assumptions.
We train our model on stellar atmospheric parameters from the LAMOST survey, which provides broad coverage of different spectral types.
We model the \Gaia XP spectra with all of their covariances, augmented by 2MASS and WISE photometry that greatly reduces degeneracies between stellar parameters, yielding more precise determinations of temperature and dust reddening.
Taken together, our approach overcomes a number of important limitations that the astrophysical parameters released in \Gaia DR3 faced, and exploits the full information content of the data. We provide the resulting catalog of stellar atmospheric parameters, revised parallaxes and extinction estimates, with all their uncertainties. The modeling procedure also produces an estimate of the optical extinction curve at the spectral resolution of the XP spectra ($R \sim 20-100$), which agrees reasonably well with the ${R(V) = 3.1}$ CCM model. Remaining limitations that will be addressed in future work are that the model assumes a universal extinction law, ignores binary stars and does not cover all parts of the Hertzsprung-Russell Diagram (\textit{e.g.}, white dwarfs).
\end{abstract}

    \section{Introduction}

\Gaia Data Release 3 (GDR3, \citealt{gdr3}) includes over 220 million flux-calibrated, low-resolution, optical stellar spectra, which provide a unique opportunity to map the properties of stars and dust throughout a large volume of the Milky Way. These low-resolution spectra are measured by two instruments, the ``Blue Photometer'' (BP), which covers the wavelength range 330-680~nm, and the ``Red Photometer'' (RP), which covers the range 640-1050~nm \citep{gdr3bprpext,grd3bprp}. Together, the BP/RP spectra (heareafter, ``XP spectra'') contain approximately 110 effective resolution elements, corresponding to a resolution of $R \sim 50 - 160$. A comparison to the largest ground-based stellar spectral survey (as of 2022), LAMOST, demonstrates the scale of the GDR3 XP spectral library. There are more than 20 times as many stars with GDR3 XP spectra as with LAMOST spectra, but each XP spectrum has approximately $1/20^{\mathrm{th}}$ the resolution of a LAMOST spectrum. The \Gaia XP spectra thus provide a unique opportunity to determine the properties of a large number of stars, but also require different modeling techniques than higher-resolution spectra.



There are a number of approaches that one might take to determine stellar parameters from the relatively low-resolution \Gaia XP spectra.
One method is to use \textit{ab initio} physical models that predict the spectrum of a star, based on its fundamental properties. For example, \cite{gspphot} uses isochrone models \citep{tang_2014_PARSEC, chen_2015_PARSEC, Pastorelli_2020_PARSEC}, four theoretical spectral atmospheric models, namely MARCS \citep{MARCS_2008}, PHOENIX \citep{Phoenix_2005}, A stars \citep{A_stars_2004} and OB \citep{OB_2008}, and the mean extinction law given by \citet{fitz99}, to fit the observed XP spectra. This approach is highly sensitive to inaccuracies in the underlying stellar models. Although the \textit{ab initio} methods could deliver precise and accurate stellar parameters if the spectral lines were resolved (and correctly modeled), the low resolution of XP spectra makes it challenging to obtain information from individual spectral lines. At the same time, because the XP spectra are flux-calibrated, their overall shape contains information about temperature and extinction. Therefore, the overall shape of the XP spectra must play a large role in constraining stellar parameters.

A second method is to learn an empirical forward model of XP spectra, using a subset of stars which have counterparts in higher-resolution spectroscopic surveys that are independent of \Gaia, and which thus have precisely determined stellar parameters. Based on this subset, a model can be built that predicts the XP spectrum as a function of the stellar parameters. We can then apply this model to all 220 million spectra, in order to infer stellar types, distances and extinctions, using standard Bayesian forward modeling techniques. This approach is attractive for a number of reasons. It is relatively interpretable, as it makes use of forward models that predict what the data \textit{should} look like, allowing exploration of residuals and discovery of new systematics and explanatory variables. In addition, this method can cope with missing data (by setting the corresponding uncertainties to infinity) and should degrade gracefully as observational uncertainties increase. In this paper, we adopt the empirical forward-modeling approach.

A third method is to train a machine-learning model to directly predict stellar parameters from XP spectra (\textit{i.e.}, supervised learning). This is similar to the second method, in that it also leverages a small subset of stars with higher-resolution spectra measured by other surveys \citep[e.g.][]{Rix2022,Andrae2023}. The key difference, however, is that supervised learning does not make use of forward modeling, but instead directly finds features in the observed spectra that are indicative of the stellar parameters. This direct machine-learning approach should degrade more rapidly in the low-signal-to-noise regime, as it does not make full use of the available measurement uncertainties. However, a model trained in this manner may also learn to correctly ignore features in the spectrum that are irrelevant to the parameters of interest (such as systematic errors and spectral features that depend on unmodeled stellar parameters). In contrast, forward modeling approaches must explicitly model all relevant parameters that have a significant effect on the observed spectra (or at the very least, introduce error terms to account for them). This downside of forward modeling may also be viewed as a strength, as it reveals the signatures of systematics and unmodeled variables in the data. Both the forward-modeling and supervised learning approach thus have merit.

Approximately 1\% of stars (or $\sim 2\times10^6$) with GDR3 XP spectra have high-quality $R \sim 1800$ measured spectra in LAMOST DR8 \citep{vaclamost} or the ``Hot Payne'' catalog \citep{hotpayne_maosheng}, and thus have well determined stellar atmospheric parameters. These stars comprehensively cover the parameter space of main sequence and giant branch, and thus allow us to build a model that predicts the XP spectrum for stars of a wide range of types. We also crossmatch this training dataset with near-infrared photometry from 2MASS \citep[$J$, $H$, and $Ks$ from ][]{2mass} and WISE \citep[$W1$ and $W2$ from][]{unwise}. Near-infrared photometry improves the determination of the stellar parameters, by helping to break the degeneracy between $\teff$ and extinction, both of which affect the overall slope of the stellar spectrum.

Our model maps \textbf{stellar parameters}, which contain the atmospheric parameters $\Theta \equiv (\teff, \logg, \feh)$, parallax ($\varpi$) and a scalar proportional to extinction ($E$), to the predicted spectrum:
\begin{equation}
    f_{\mathrm{pred}}(\lambda\ |\ \Theta,\varpi,E)
    = f_{\mathrm{abs}}(\lambda\ |\ \Theta)
      \varpi^2 \exp\left[-E\,R(\lambda)\right]
    \, ,
\end{equation}
where $f_{\mathrm{abs}}$ is the ``absolute flux'' of a star at 1~kpc, as a function of wavelength ($\lambda$). We represent the mapping from $\Theta$ to $f_{\mathrm{abs}}$ as a neural network, and the extinction curve $R\left(\lambda\right)$ as a vector (with one entry per wavelength). The structure of our model encodes certain reasonable assumptions:
\begin{enumerate}
    \item Flux falls with the square of distance.
    \item Dust imposes a wavelength-dependent optical depth.
    \item In the absence of dust (and at a standard distance), the stellar spectrum is purely a function of stellar atmospheric parameters.
\end{enumerate}
Strictly speaking, the intrinsic stellar parameters should be represented by the initial mass, age and elemental abundances of the stars, which can then be mapped to stellar atmospheric parameters using models of stellar evolution. However, stellar atmospheric parameters can be more directly determined from observed stellar spectra (without assuming a stellar evolutionary model). We assume that there exists a one-to-one mapping between the initial mass, age and elemental abundances and the atmospheric parameters $(\teff, \logg, \feh)$, and use the latter to represent stellar type. This assumption is valid across most of the Hertzsprung-Russell Diagram, and has been adopted by \cite{sch16} and \cite{ddm}. Two limitations of our present model are that we assume a universal dust extinction curve (\textit{i.e.}, extinction is always proportional to a universal function, $R\left(\lambda\right)$), and that we do not model $\left[\alpha/\mathrm{Fe}\right]$. In fact, our parameter $\feh$ should not be viewed strictly as a measure of iron abundance, but rather as a measure of whichever metals are most apparent in XP spectra (including, potentially, of $\alpha$ elements). An additional limitation of our method is that we treat every source as a single star, although a non-negligible proportion of them must be binaries. We discuss these limitations -- and possible ways to address them -- in Section~\ref{sec:discussion}.

We build our model in an auto-differentiable framework, which allows us to easily calculate gradients of our model with respect to any of the parameters. This has a number of important advantages:
\begin{enumerate}
    \item We can use gradient descent to maximize the likelihood of the observed data, or the posterior density of our model parameters and individual stellar types.
    \item We can propagate measurement uncertainties through our model to obtain uncertainty estimates on our stellar parameters.
\end{enumerate}
We optimize the model by maximizing the likelihood of the observed GDR3 XP spectra and 2MASS/WISE near-infrared photometry. We also update the estimates of the individual stellar parameters, using both the likelihood and prior distributions of $\Theta$ from LAMOST, $\varpi$ from GDR3, and $E$ from Bayestar19 \citep{bayestar19}. We iteratively switch between updating the model coefficients and optimizing the individual stellar parameters until both converge. The extinction curve $R(\lambda)$, as part of the model, is also optimized. We find that the extinction curve is smooth, as a function of $\lambda$, even if no constraints on smoothness are applied, and that the extinction curve is consistent with other widely-used models, such as the model by \cite{ccm}. We note that our modeling approach in this paper is similar to the approach taken by \citet{ddm} to model stellar photometry, in that we directly learn an auto-differentiable, empirically driven forward model from stars with spectroscopic measurements.
 
After our model is trained on the training set, which composes of the 1\% of stars having LAMOST DR8 and ``Hot Payne'' counterparts, we apply the model to fit the rest of the GDR3 XP spectra, in order to constrain their stellar types, distances and extinctions. We also investigate the resulting preliminary 3D distribution of stellar types, as a ``forerunner'' of a next-generation 3D dust map. From the residual of the flux at different wavelengths, we notice the signatures of the variation of the extinction law.   
 
In this work, we obtain stellar type estimates from LAMOST. However, our approach would also work with stellar type estimates from further spectroscopic surveys. For example, our model can be combined with the incoming spectroscopic data from the SDSS-V Milky Way Mapper (MWM), which covers stars deeper into the disk \citep{MWM}. MWM will also provide data with better resolution in the infrared, which will significantly improve the precision of determination of stellar types \citep{MWM}. Moreover, in the future version of our model, we will introduce additional parameter representing the variation of extinction law, which helps further exploring the physical and the Milky Way structural influence of the dust extinction, as well as further improve the precision of 3D dust maps in the Milky Way.

This paper is organized as follows. In Section~\ref{sec:data}, we discuss the spectroscopic and photometric datasets and the crossmatching. We also discuss the processing of the error of these observations. In Section~\ref{sec:method}, we explain the structure of our forward model of XP spectra, and discuss how we train it using XP spectra with matched LAMOST observations. In Section~\ref{sec:optimization}, we discuss how we use the trained forward model to infer stellar parameters for all 220 million sources with XP spectra. In Section~\ref{sec:results}, we present our trained model, our catalog of inferred stellar parameters, and a preliminary three-dimensional dust map based on our stellar reddening and distance estimates. Finally, in Section~\ref{sec:discussion}, we discuss possible uses and further extensions of our model and stellar parameter inferences.

    \section{Data}
\label{sec:data}

We infer stellar parameters using three sources of spectro-photometric data (\Gaia XP, 2MASS and unWISE), as well as \Gaia parallaxes. The inclusion of near-infrared photometry from 2MASS and WISE helps us disentangle stellar temperature and extinction, by anchoring our model at longer wavelengths, where extinction has a far smaller effect. In order to train our stellar model, we additionally make use of stellar atmospheric parameter estimates based on LAMOST spectroscopy. We describe each source of data in greater detail below.

\subsection{\Gaia XP spectra}
\label{sec:xp-spectra}

\Gaia is a satellite-based observatory launched in 2013 by the European Space Agency (ESA), aimed at providing 6-D (sky position, parallax, proper motion and radial velocity) astrometry measurements for objects in the Milky Way and Local Group \citep{gaia_mission}. \Gaia Data Release 3 (GDR3) provides positions and parallaxes of $1.46\times10^9$ sources \citep{gdr3}. On the \Gaia satellite, there are two spectrophotometers, with the ``Blue Photometer'' (BP) covering 330-680~nm and the ``Blue Photometer'' (RP) covering 640-1050~nm \citep{grd3bprp}. As of GDR3, BP/RP spectra (hereafter ``XP spectra'') of $\sim220$ million stars have been published \citep{gdr3,internalcali}.

\Gaia XP spectra are not natively reported as pixelized wavelength-space measurements. Instead, BP and RP spectra are each projected onto 55 orthonormal Hermite functions, and reported as a set of coefficients. Therefore, the XP spectrum of each star is represented by a 110-dimensional coefficient vector. The reason for such a representation is that the observation of each star is the combination of a series of epochs. In different epochs, the instrumental influence varies with time, the focal plane position, the detector used, the field of view, and other factors. These differences are the integral transforms of the PSF, and therefore it is mathematically easier to work with a series of continuous orthogonal basis functions \citep{internalcali,gdr3bprpext}. However, for our application, it is preferable to transfer these coefficients back to wavelength space for two major reasons: 
\begin{enumerate}
    \item The dust extinction effect is linear in magnitude space, and it would be complicated to parameterize the extinction in the space of Hermite-function coefficients.
    \item The spectra are noisy at the edges of BP and RP, including in their overlap region. In Hermite space, the noisy part of the spectrum is encoded by a wide range of coefficients. It is therefore impossible to remove noisy data in the edges of the BP and RP wavelength ranges by excluding certain coefficients.
\end{enumerate}

We use \texttt{GaiaXPy} package\footnote{GaiaXPy is described at \url{https://gaia-dpci.github.io/GaiaXPy-website/}, and version 1.1.4 can be found at \url{https://doi.org/10.5281/zenodo.6674521}} to convert the XP coefficients to sampled (\textit{i.e.}, wavelength-space) spectra. Cutting out the noisy edges of the XP spectral range results in the loss of degrees of freedom. We sample the spectra from 392-992~nm, in increments of 10~nm. We find that this is the finest sampling that reliably ensures that the covariance matrices of the sampled spectra remain positive-definite. Because the flux in our representation is sampled at discrete wavelengths, we refer to it as the ``\textbf{flux vector}.''

In order to accurately model XP spectra, it is also critical to include information about the covariances between the observed fluxes at different wavelengths, which are in general non-negligible. We use \texttt{GaiaXPy} to transfer the covariance matrices of the 110 basis-function coefficients to the sampled space.\footnote{Note that we do not scale the BP and RP covariance matrices by \texttt{bp\_standard\_deviation} and \texttt{rp\_standard\_deviation}, respectively, due to an oversight during the preparation of the data. These factors typically change the uncertainties by only a few percent, though the corrections can be larger for a small subset of stars.} We denote the resulting sample-space covariance matrix as $C_{\vec{f}_{\text{gaia}}}$. To prevent individual wavelengths with very small reported uncertainties from dominating our results, we inflate the covariance matrix by introducing a lower limit on the individual flux uncertainties. We additionally add in zero-point uncertainties to the covariance matrices. The final covariance matrix that we use, $C_{\vec{f}_{\text{obs}}}$, is given by
\begin{align}
    C_{\vec{f}_{\text{obs}}}
    &=
    C_{\vec{f}_{\text{gaia}}}
    \!\!\! + \,
    \operatorname{diag} \left( 0.005 \vec{f}_{\text{gaia}} \right)^2
    + 0.005^2 \vec{f}_{\text{gaia}} \vec{f}_{\text{gaia}}^{\, T}
    \notag \\
    &\hspace{1cm}
    + 0.001^2 \vec{f}_{\text{BP}} \vec{f}_{\text{BP}}^{\, T}
    + 0.001^2 \vec{f}_{\text{RP}} \vec{f}_{\text{RP}}^{\, T}
    \, ,
    \label{eqn:cov-obs}
\end{align}
where $\vec{f}_{\text{gaia}}$ is the sampled flux, and $\vec{f}_{\text{BP}}$ and $\vec{f}_{\text{RP}}$ are zero-padded vectors containing the sampled BP and RP spectra, respectively. The first term is the sample-space covariance matrix, calculated directly from \Gaia's reported coefficient-space covariance matrix. The second term allows the flux at each wavelength to vary independently by $0.5\%$. The third term allows the zero-point of the overall luminosity of the entire spectrum to vary by $0.5\%$. The fourth and fifth terms allow the zero-points of the BP and RP sides of the spectrum to independently vary by $0.1\%$, in order to capture possible errors in the relative zero-point calibration of BP vs. RP.

The inflation of the flux covariance matrix ensures reasonable error bars for each element in the flux vector, but it cannot guarantee that the covariance matrix as a whole is well behaved (for example, that the condition number is small). We therefore diagonalize each covariance matrix using its eigendecomposition:\footnote{We use \texttt{numpy.linalg.eigh} for the eigendecomposition of the covariance matrices, taking advantage of the fact that they are Hermitian.}
\begin{equation}
    C_{\vec{f}_{\text{obs}}} = U D U^T \, ,
    \label{eqn:cov-decomp}
\end{equation}
where $U$ is an orthonormal matrix and $D$ is a diagonal matrix. There are sometimes extremely small (or even more rarely, negative) values in $D$, which means that the predicted flux vectors are extremely strongly constrained along the corresponding eigenvectors. These ``constraints'' are far stronger than those one would expect from the typical scale of the flux uncertainties, and can lead to practical difficulties during training. We limit the strength of these problematic constraints by setting the minimum allowable value along the diagonal of the $D$ matrix to $10^{-9}$. We denote the modified $D$ matrix as $\hat{D}$, and let $L \equiv \hat{D}^{-1/2} U^T$, where $\hat{D}^{-1/2}$ is well defined because $\hat{D}$ is diagonal. We can then calculate the $\chi^2$ between the predicted flux and observed flux as
\begin{align}
    \chi^2
    &=
      \Delta \vec{f}^{\, T}
      C_{\vec{f}_{\text{obs}}}^{-1}
      \Delta \vec{f}
    =
    \left|
    L \, \Delta \vec{f} \,
    \right|^2
    \, ,
    \label{eqn:chi2}
\end{align}
where $\Delta \vec{f} \equiv \vec{f}_{\text{pred}}-\vec{f}_{\text{obs}}$ is the residual between the predicted and observed flux.

\begin{figure}
    \centering
    \includegraphics[width=1.0\linewidth]{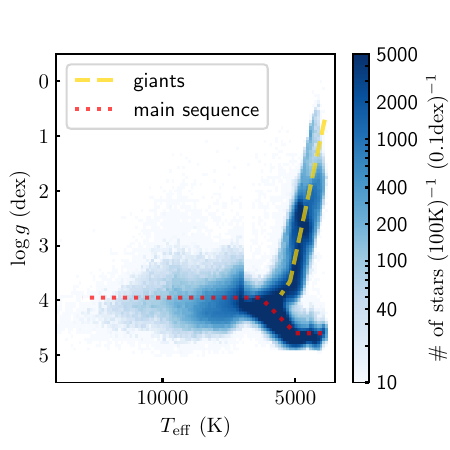}
    \caption{Kiel diagram ($\logg\  vs.\  \teff $) of our matched GDR3~XP and LAMOST catalog, used for training and validating our stellar model. The matched catalog contains 2,575,354 stars, covering much -- but not all - of stellar parameter space. The sharp feature at $\teff = 7000\,\mathrm{K}$ is the point at which we transition from the standard DR8 stellar atmospheric parameter estimates to the ``Hot Payne'' \citep{hotpayne_maosheng}, which provides more accurate temperatures for hot stars. We overplot piecewise-linear main-sequence and giant-branch tracks (See Eqs.~\ref{eqn:logg-ms} and \ref{eqn:teff-giants}), which we later use when visualizing our trained stellar models.}
    \label{Fig:LAMOST-crossmatch}
\end{figure}

\subsection{LAMOST}
\label{sec:lamost}

The Large Sky Area Multi-Object Fiber Spectroscopic Telescope (LAMOST) is a ground-based telescope which has observed over $\sim10$ optical stellar spectra in the Northern Hemisphere with a resolution $R \sim 1800$ and with the limiting magnitude of $r<19$ \citep{cui_lamost,zhao_lamost}. 
In this work, we use the AFGK catalog from LAMOST Data Release 8 (DR8),\footnote{LAMOST DR8: \url{http://www.lamost.org/dr8/v2.0/}} which contains atmospheric parameters of over 6 million stars. The typical errors of the stellar parameters in this catalog are $\sim40\,\mathrm{K}$ for $\teff$, 0.06~dex for $\logg$, and 0.04~dex for $\feh$. For hot stars ($\teff \gtrsim 7500 \,\mathrm{K}$), we obtain stellar atmospheric parameters from the ``Hot Payne'' \citep{hotpayne_maosheng}, which applies the method, ``The Payne'' \citep{thepayne_ting} to determine the atmospheric parameters of $\sim330,000$ O-, B- and A-type stars in LAMOST Data Release 6. The standard LAMOST stellar parameter pipeline has difficulty modeling the Balmer lines of stars with $8,000\,\mathrm{K} \lesssim \teff \lesssim 12,000\,\mathrm{K}$, leading to inaccurate temperature estimates in this range. The Hot Payne produces more accurate temperature estimates for stars in this regime. However, as can be seen in Fig.~\ref{Fig:LAMOST-crossmatch}, the distribution of $\logg$ in our combined catalog changes sharply at 7000~K, where we transition from the standard LAMOST catalog to the Hot Payne catalog. We leave the problem of spanning these two temperature regimes more seamlessly to future work.


We crossmatch LAMOST with GDR3 by searching the closest star, out to a maximum angular separation of $0.25"$. Fig.~\ref{Fig:LAMOST-crossmatch} shows a Kiel diagram of the resulting cross-matched catalog. The matched stars cover a wide area of stellar atmospheric parameter space, including the main sequence from ${\sim 4,000 - 12,000 \,\mathrm{K}}$, and the giant branch up to $\logg \sim 0.5$. Certain stellar types and stellar remnants, such as white dwarfs and subdwarfs, are not covered by our matched catalog. In addition, some regions of parameter space, such as the horizontal branch and stars with $\teff \gtrsim 15,000\,\mathrm{K}$, have only sparse coverage. Although the overall number of matched sources ($\sim 2$ million) comprises only $\sim1\%$ of the \Gaia XP catalog, it is still sufficient to train a model of XP spectra over a wide range of stellar parameter space.


\subsection{2MASS}
\label{sec:2MASS}

2MASS is a near-infrared, all-sky imaging survey, which obtained photometry for $\sim4\times10^8$ stars in three bands, namely $J$, $H$ and $\mathrm{K_s}$ \citep{2mass}. The effective wavelengths of the bands are $J\ (1.25 \mu \mathrm{m}), H\ (1.65 \mu \mathrm{m})$, and $K_s (2.16 \mu \mathrm{m})$, respectively. The 10~$\sigma$ limiting sensitivities for these bands are $J<15.8$, $H<15.1$ and $K_s<15.8$. The typical uncertainty for bright sources ($K_s<13$) is 0.03 mag, and the calibration offsets are < 0.02 mag.

We use the crossmatched 2MASS data from the \Gaia Archive \citep{cm2massbestneighbour,cm2massjoin,gaiaofficial}. See Appendix~\ref{sec:gaia-archive-queries} for our full ADQL query.

\begin{figure*}
    \centering
    \includegraphics[width=1.0\linewidth]{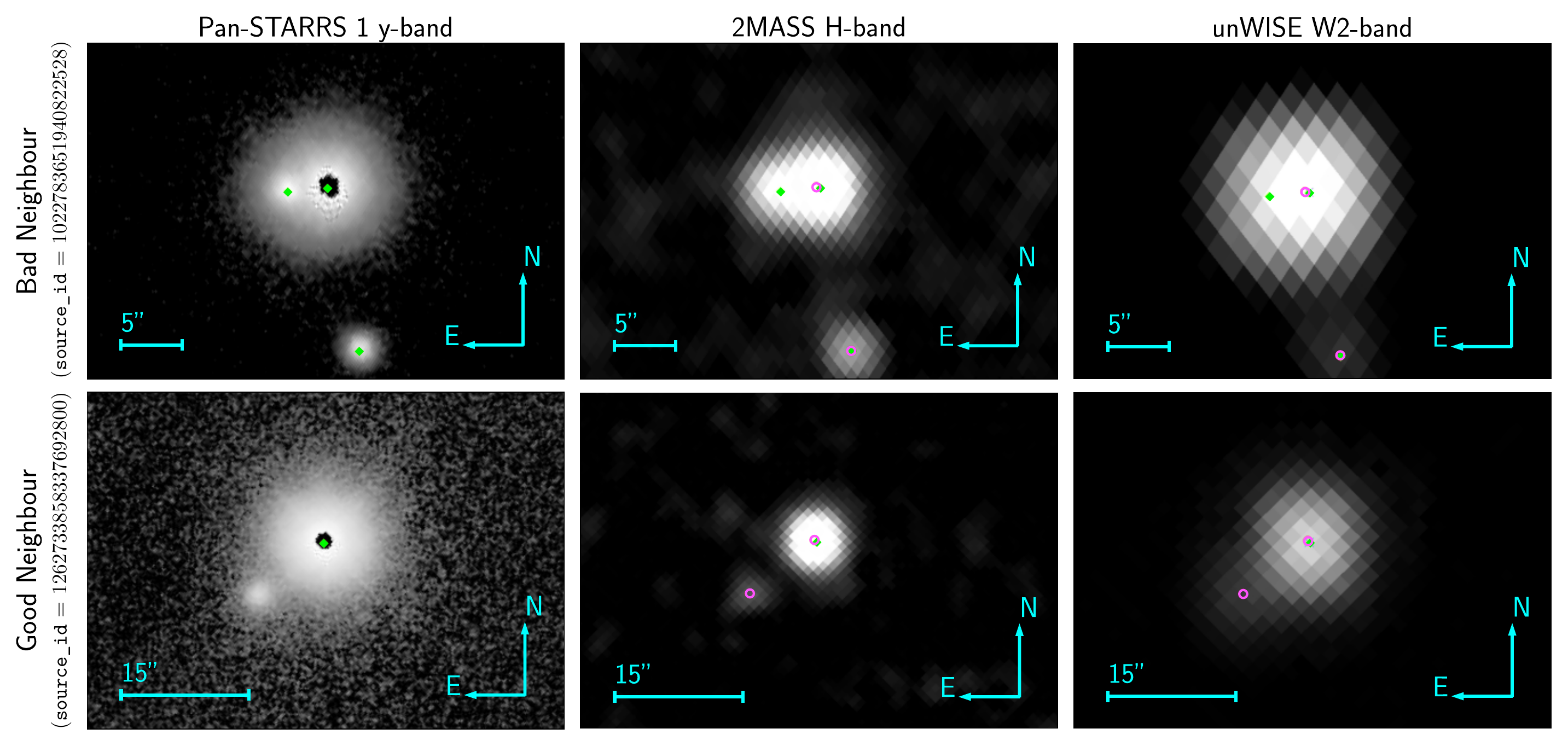}
    \caption{Example of a source (centered in each image) with a problematic (top panels) and an unproblematic neighbor (bottom panels). The left panels show PS1 $y$-band images, the center panels show 2MASS $H$-band images, and the right panels show unWISE $W2$-band images. The green diamonds (\textcolor[rgb]{0,1,0}{$\blacklozenge$}) mark the locations of sources with XP spectra in GDR3. In the middle and right panels, the violet circles (\textcolor[rgb]{1,0.329,1}{$\boldsymbol{\circ}$}) mark the locations of point sources included in the 2MASS and unWISE catalogs, respectively. 2MASS and WISE do not detect the problematic neighbor, due to their lower resolution (approximately 2" and 6", respectively) compared to \Gaia. The source with the problematic neighbor has a relatively high \texttt{norm\_dg} value of -5.98, indicating that it has a close, bright neighbor. This source fails our cut for including unWISE measurements, but barely passes our cut for using 2MASS measurements. The bottom panels show a source with an unproblematic neighbor, which is detected in both 2MASS and unWISE. As this source has a \texttt{norm\_dg} value of -19.4, indicating that it is relatively isolated, we use both its 2MASS and unWISE measurements.}
    \label{fig:unresolve}
\end{figure*}

2MASS used three $256\times256$ pixel arrays, with a pixel scale of $2^{\prime \prime} \times 2^{\prime \prime}$, which is much larger than that of \Gaia CCD pixels ($\sim177 \, \mathrm{mas} \times \ 59 \, \mathrm{mas}$, \citealt{gaiasummary}). Therefore, a star clearly resolved by \Gaia could be severely contaminated by its neighbors in 2MASS, as shown in Fig.~\ref{fig:unresolve}. The two stars at $(\alpha_{J2000},\delta_{J2000})$= (09:17:13.8, 53:25:09.6) are resolved by GDR3, and fulfill the standard of GDR3 to be spectroscopically analysed, but they cannot be resolved by 2MASS because their angular separation is $\sim3.2^{\prime\prime}$. In such cases, we expect a positive bias in the observed $J$, $H$ and $K_s$ fluxes, compared with the predicted fluxes given by our model. To solve this, we disregard 2MASS observations (by setting their variance to infinity) if $\mathtt{norm\_dg} > -5$. The parameter \texttt{norm\_dg} is defined as $max\{\frac{\Delta G_i}{\mathrm{mag}}-\frac{\theta_i}{\operatorname{arcsec}}\}$, where $\Delta G_i$ is the difference of G-band magnitude between the target star and its {i-th} neighbor with an angular separation $<30^{\prime\prime}$, and $\theta_i$ is the angular separation between them \citep{gedr3fidelity}. Therefore, \texttt{norm\_dg} is a measure of crowding, with larger values indicating the presence of closer, brighter neighbors.

2MASS photometry is natively provided in Vega magnitudes. In order to put 2MASS photometry on a scale that is easily comparable to \Gaia XP spectral fluxes, we convert it to spectral flux using an assumed central wavelength and AB-Vega magnitude offset:
\begin{align}
    f_{\lambda} = \left(3631\,\mathrm{Jy}\right) c \lambda_0^{-2} 10^{-0.4 \left(m + \Delta m \right)} \, .
    \label{eqn:vega-to-spectral-flux}
\end{align}
Where $m$ is the reported Vega magnitude; $c$ is the speed of light; the AB-Vega offset $\Delta m = 0.91$, 1.39 and 1.85~mag for $J$, $H$ and $K_s$, respectively \citep{Blanton2005ZeroPoints}; and the central wavelength $\lambda_0 = 1.235$, 1.662 and 2.159~$\mu$m for $J$, $H$ and $K_s$, respectively. This conversion is only approximate, as a correct conversion between broad-band photometry and spectral flux requires a knowledge of the shape of the source spectrum. Nevertheless, this conversion is mathematically well defined, can be inverted to recover the broad-band photometry, and allows a rough comparison with XP spectral fluxes.

\subsection{unWISE}
\label{sec:WISE}

The WISE mission has observed the entire sky in four bands, $W1$ $(3.4 \mu m)$, $W2$ $(4.6 \mu m)$, $W3$ $(12 \mu m)$ and $W4$ $(22 \mu m)$, with a space-based 40~cm telescope (``WISE cryogenic'', \citealt{wiseofficial}). Because the HgCdTe detector arrays used by $W1$ and $W2$ remain functional without solid hydrogen as coolant, more data in these two bands was accumulated in search of near-Earth objects before the hibernation in 2011 (``NEOWISE'', \citealt{neowise11}). Yet more $W1$ and $W2$ observations have been collected since the reactivation of NEOWISE in 2013 (``NEOWISER'', \citealt{neowise14}). \citet{unwise} co-added ``WISE cryogenic'', ``NEOWISE'' and ``NEOWISER'' to build the ``unWISE'' catalog, which is deeper than each individual catalog it used. We make use of the unWISE ``neo6'' catalog, which is based on WISE exposures captured through 13 December 2019 \citep{Meisner2021unwiseneo6}. The unWISE catalog behaves better in the Galactic disk, as it uses the ``crowdsource'' pipeline, which was developed to handle the extremely crowded fields observed by the Dark Energy Camera Plane Survey \citep{decapplane}.

The FWHM of the point spread function (PSF) of WISE is $6''.1$ for $W1$ and $6''.4$ for $W2$, which is several times larger than GDR3. For this reason, we disregard unWISE photometry if ${\mathtt{norm\_dg} > -10}$, indicating the presence of nearby, bright neighbors.

As with 2MASS, we convert unWISE magnitudes from their native Vega system to spectra flux using Eq.~\eqref{eqn:vega-to-spectral-flux}. We assume central wavelengths of $\lambda_0 = 3.3526$ and 4.6028~$\mu$m, and and AB offsets of $m_0$ 2.699 and 3.339~mag for $W1$ and $W2$, respectively.\footnote{WISE Vega-AB magnitude offsets are given by \url{https://wise2.ipac.caltech.edu/docs/release/allsky/expsup/sec4_4h.html\#conv2ab}.}

\subsection{Training and validation datasets}
\label{sec:training-validation-datasets}

We train our model on our cross-matched \Gaia~XP--LAMOST catalog, with \Gaia parallaxes and photometry from 2MASS and unWISE. We impose a number of additional quality cuts on our cross-matched catalog:
\begin{itemize}
    \item LAMOST SNR of greater than $20$ in $g$-, $r$-, and $i$-bands.
    \item LAMOST uncertainties of less than 500~K in $\teff$, and less than 0.5~dex in $\feh$ and $\logg$.
    \item Well constrained parallaxes: ${\mathtt{parallax\_over\_error} > 3}$.
    \item Reliable \Gaia astrometry: ${\mathtt{fidelity\_v2} > 0.5}$.
    \item Low BP/RP flux excess: ${\mathtt{bp\_rp\_flux\_excess} < 1.3}$.
\end{itemize}
After these quality cuts, our cross-matched catalog contains a total of 2,575,354 sources. We split our dataset into a training set (80\% of sources) and a validation set (20\% of sources). The training dataset is used to train a model of stellar flux and dust extinction, as a function of wavelength, while the validation set is used to evaluate the performance of the trained model.
	
    \section{Method}
\label{sec:method}

\begin{figure*}
    \centering
    \includegraphics[width=1.0\linewidth]{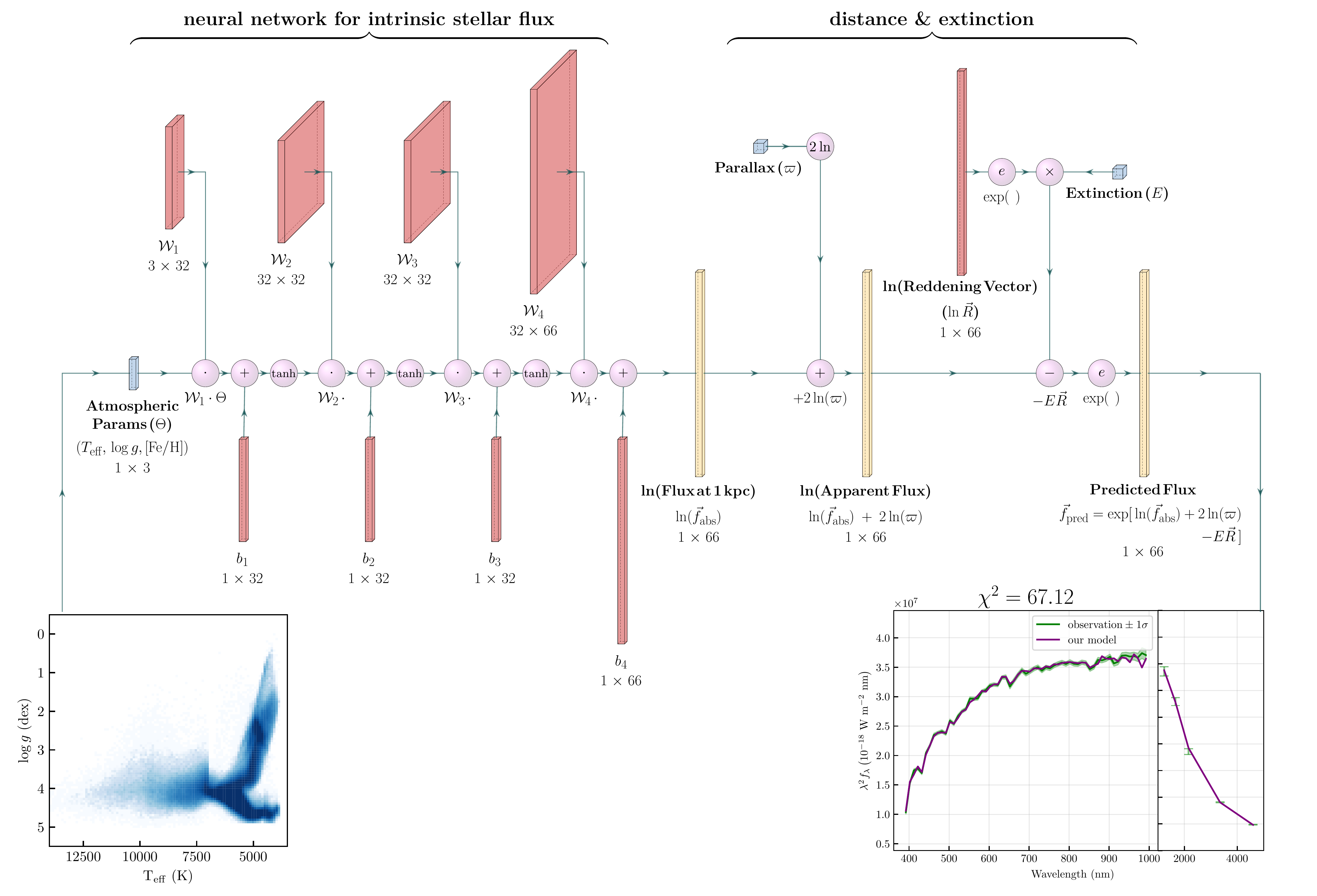}
    \caption{The structure of our stellar flux model. Individual stellar parameters are represented in the light blue blocks, $\Theta\equiv (\teff, \logg$, [Fe/H]), the parallax $\varpi$ and the extinctions $E$. Global model parameters are represented as red blocks, and intermediate calculations and outputs are represented in yellow. The overall structure of the model is chosen to only use neural networks in the aspects where simple \emph{ab initio} physical models fall short. The stellar atmospheric parameters ($\Theta$) are fed into a neural network consisting of three 16-neuron dense layers, with weights $\mathcal{W}$ and biases $b$. The neural network outputs the natural logarithm of the unreddened absolute flux observed at 1~kpc, which is a 66-dimensional vector (61 wavelength samples from XP with an interval of 10~nm, plus $J$, $H$, $\rm K_s$ from 2MASS and $W1$ and $W2$ bands from unWISE). The inverse-square law decay in observed flux is accounted for with the factor $\varpi^2$, and the extinction is represented by the factor of $\exp(-E\vec{R})$, where $\vec{R}$ is the relative amount of extinction at each wavelength and $E$ is a scalar that represents the overall amount of extinction for the given star. The final output is the predicted flux, $\vec{f}_{\rm pred}=\vec{f}_{\rm abs}\varpi^2\exp(-E\vec{R})$. This prediction is compared with the observation, $\vec{f}_{\rm obs}$, to calculate $\chi^2$. The loss function consists of the $\chi^2$, the likelihoods of ($\Theta$, $\varpi$, $E$) from outside observations (LAMOST, \Gaia and Bayestar19, respectively), and the L2 regularization of the neural network weights ($\mathcal{W}$). In the training process, we first fix ($\Theta$, $\varpi$, $E$), the blue blocks in the picture, and find the optimal global parameters, ($\mathcal{W}$, $b$, $\vec{R}$), that minimize the loss function. Then we fix ($\mathcal{W}$, $b$, $\vec{R}$), the red blocks, and optimize the stellar parameters ($\Theta$, $\varpi$, $E$). We repeatedly switch between updating the red blocks and optimizing blue blocks, until the loss function stops decreasing. 
    }
    \label{fig:model-structure}
\end{figure*}

We build a forward model that maps from stellar type, extinction and parallax to the expected XP spectrum and near-infrared 2MASS and WISE photometry. We build our model in the auto-differentiable TensorFlow~2 framework \citep{tensorflow2015-whitepaper}, which allows us to optimize both the model's internal parameters and to fit individual stellar parameters using gradient descent methods. We train our forward model using atmospheric parameters from LAMOST, parallaxes from \Gaia DR3 \citep{gdr3}, and reddenings from the 3D dust map Bayestar19 \citep{bayestar19}. We then apply our model to all stars with \Gaia XP spectra, to determine their types, distances and extinctions.


We map stellar atmospheric parameters, $\Theta = (\teff, \logg, \feh)$, to the unreddened spectral profile at 1~kpc, which we call the ``absolute flux'' and denote by $\vec{f}_{\mathrm{abs}}(\Theta)$:
\begin{equation}
    \Theta \rightarrow \vec{f}_{\mathrm{abs}}(\Theta) \, .
\end{equation}
We implement this model as a simple feed-forward neural network, which takes $\Theta$ as an input and outputs a vector, with each entry representing the natural log of the absolute flux at a different wavelength. To obtain the unreddened flux at the distance of the star, we multiply by $\varpi^2$, the square of the parallax (note that this is a model parameter, and not the observed noisy parallax). We assume that the effect of extinction is to multiply the absolute flux by
\begin{equation}
    \exp(-E\vec{R}) \, ,
\end{equation}
where $E$ is a scalar measurement of the amount of extinction in front of the star along the line of sight. The vector $\vec{R}$ is shared by all stars, and represents the relative amount of extinction at each wavelength. We ignore extinction effects that would be caused by the finite width of our bandpasses (see Appendix~B of \citealt{ddm} for a discussion of such effects), because XP spectral elements are much narrower than typical photometric bands. For 2MASS and WISE photometry, this will introduce somewhat larger fractional errors in extinction, but this effect is mitigated by the fact that extinction is smaller in the near-infrared.
 
Putting all the piece of our model together, the predicted flux is given by
\begin{align}
\label{eqn:full_model}
	\vec{f}_{\mathrm{pred}}
	&=
	\vec{f}_{\mathrm{abs}}(\Theta,\mathcal{W},b)
	\varpi^2
	\exp \big(-E \vec{R}\big) \, ,
\end{align}
where $\mathcal{W}$ and $b$ represent all trainable neural-network weights and biases, respectively, in the absolute flux model. The structure of our model is shown in Fig.~\ref{fig:model-structure}. In total, our model has 4484 global parameters (contained in $\mathcal{W}$ and $b$, the weights and biases of the neural network, as well as the extinction curve $\vec{R}$), and an additional five parameters per star ($\teff$, $\feh$, $\logg$, $\varpi$, and $E$).

In the following, we provide a detailed description of the workings of our model, and of our training procedure. We invite readers who are primarily interested in the results we obtain using the model to inspect Fig.~\ref{fig:model-structure}, which gives an overview of the model structure, and then to proceed to Section~\ref{sec:results}.

\subsection{Bayesian formulation of model}

Before describing how we learn the parameters in the model, we will write down the posterior density of the model and individual stellar parameters, given our observations. Then, we will show how we use gradient descent to infer both the global model parameters and the individual stellar parameters.

The likelihood of a single star's observed flux $\vec{f}_{\mathrm{obs}}$ (with uncertainties described by the covariance matrix $C_f$), given our global model parameters $\mathcal{W}$, $b$ and $\vec{R}$, as well as our modeled type $\Theta$, true stellar parallax $\varpi$, and reddening $E$ is given (up to a constant) by
\begin{align}
    \ln p \left(
      \vec{f}_{\mathrm{obs}}
      \mid
      \mathcal{W}, b, \vec{R}, \Theta, \varpi, E
    \right)
    &=
    -\frac{1}{2} \underbrace{
        \Delta \vec{f}^{\, T} \, C_f^{-1} \Delta \vec{f}
    }_{\equiv \chi^2}
    \, ,
    \label{eqn:model-likelihood}
\end{align}
where $\Delta \vec{f} \equiv \vec{f}_{\mathrm{obs}} - \vec{f}_{\mathrm{pred}} \left(\Theta, \varpi, E \mid \mathcal{W}, b, \vec{R} \right)$. In practice, we calculate $\chi^2$ using our matrix decomposition of $C_f^{-1}$, as shown in Equation~\eqref{eqn:chi2}.

For each star, we also have an observed \Gaia parallax $\hat{\varpi}$ (with Gaussian uncertainty $\sigma_{\varpi}$), a LAMOST estimate of stellar type $\hat{\Theta}$ (with diagonal covariance matrix described by $\vec{\sigma}_{\Theta}$), and a Gaussian estimate of reddening which we derive from Bayestar19 (with uncertainty $\sigma_E$). We treat these as observed quantities, so that each contributes an independent Gaussian likelihood term. We additionally place priors on the individual stellar parameters. We place improper flat priors on $\varpi$, $E$, and $\Theta$, with the additional condition that $\varpi$ and $E$ must be positive (we achieve this by using $\ln \varpi$ and $\ln E$ as our variables, and including the appropriate Jacobian terms to transform our flat priors in $\varpi$ and $E$).

We also impose simple priors on the global model parameters, as follows. We place a flat prior on the logarithm of each component of $\vec{R}$, a Gaussian prior on the neural network weights $\mathcal{W}$ (which is equivalent to adding an L2 regularization penalty to the network), and a flat prior on the neural network biases $b$. As will be seen later, we set the standard deviation of the prior on $\mathcal{W}$ so that it makes a similar contribution to the posterior as one extra degree of freedom per star would be expected to make.

In full, the posterior of our model, assuming one observed star, can be written as
\begin{align}
    &
    p \left(
      \mathcal{W}, b, \vec{R}, \Theta, \varpi, E
      \mid
      f_{\mathrm{obs}},
      \hat{\Theta}, \sigma_{\Theta},
      \hat{\varpi}, \sigma_{\varpi},
      \hat{E}, \sigma_{E}, 
      C_f
    \right)
    \notag \\
    &\hspace{1cm}
    \propto
    p \left(
      \vec{f}_{\mathrm{obs}}
      \mid
      \mathcal{W}, b, \vec{R}, \Theta, \varpi, E, C_f
    \right)
    \notag \\
    &\hspace{1cm}\ \ 
    \times
    p \left(
      \hat{\varpi}
      \mid
      \varpi, \sigma_{\varpi}
    \right)
    p \left(
      \hat{E}
      \mid
      E, \sigma_{E}
    \right)
    p \left(
      \hat{\Theta}
      \mid
      \Theta, \sigma_{\Theta}
    \right)
    \notag \\
    &\hspace{1cm}\ \ 
    \times
    p\left(\varpi\right) p\left(E\right) p\left(\Theta\right)
    p\left(\mathcal{W}\right) p\left(b\right) p\big(\vec{R}\big)
    \, ,
    \label{eqn:model-posterior}
\end{align}
with a normalizing constant that is independent of the model parameters. The generalization to multiple stars is trivial: every term except for the priors on $\mathcal{W}$, $b$ and $\vec{R}$ is repeated for each star.

\subsection{Training the model}
\label{sec:training}

\begin{figure*}
    \centering
    \includegraphics[width=1.0\linewidth]{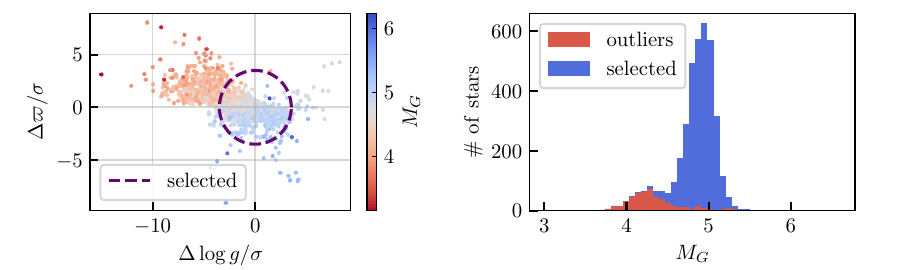}
    \caption{``Self-cleaning'' process of our approach, which generates outlier rejection flags. In essence, the approach identifies instances where the model's fundamental assumption  -- stellar parameters uniquely predict absolute spectral fluxes -- create significant tension with the data (e.g. the parallax). For the illustrative regime of low-extinction Solar-type stars (defined by Eq.~\ref{eqn:solar_type}), the left panel shows the deviation of our optimized $\varpi$ and $\logg$ from the observed values. Each star is colored by the inferred absolute magnitude, $M_G = m_G - \mu$, where $m_G$ is the apparent $G$-band magnitude, and the distance modulus is given by $\mu = 10-5\log(\varpi/1\ \mathrm{mas})$. We define $\Delta \varpi/\sigma \equiv (\varpi_\mathrm{opt}-\varpi_\mathrm{obs})/\sigma(\varpi_\mathrm{obs})$ and $\Delta \logg/\sigma \equiv (\logg_\mathrm{opt}-\logg_\mathrm{obs})/\sigma(\logg_\mathrm{obs})$. We remove stars with large residuals, $(\Delta \logg/\sigma)^2+(\Delta \varpi/\sigma)^2>3.5^2$. The right panel shows a histogram of the absolute $G$-band magnitude of same population of stars, with the outliers identified in the left panel colored in red. The outliers form a distinct population with absolute magnitudes that are $\sim 0.75\,\mathrm{mag}$ brighter (corresponding to twice the luminosity) than the typical star with the same LAMOST atmospheric parameters. The outliers are thus likely dominated by near-equal-mass binaries.}
    \label{fig:selfcleaning}
\end{figure*}

\begin{table*}
    \begin{tabular}{|c|c|c|c|c|c|c|}
        \hline
        Phase & Only HQ data & Initial LR of ($\mathcal{W}$, $b$, $\vec{R}$) & Initial LR of ($\Theta$, $\varpi$, $E$) & Optimizer \\ 
        \hline
        1& $\checkmark$ & $10^{-3}$ & $-$ & SGD \\ 
        \hline
        2& $\checkmark$ & $10^{-5}$ & $10^{-4}$ & SGD\\ 
        \hline
        3&  & $-$ & $10^{-4}$ & SGD\\ 
        \hline
        4&\multicolumn{4}{c}{{\color{gray}\rule[2pt]{10em}{0.1pt}} \hspace{1em} Self-cleaning \hspace{1em} {\color{gray}\rule[2pt]{10em}{0.1pt}}} \\
        \hline
        5&  & $10^{-7}$ & $10^{-5}$ & SGD \\ 
        \hline
        6&  & $-$ & $10^{-2}$ & Adam \\ 
        \hline
    \end{tabular} 
    \caption{Training procedure described in Section~\ref{sec:training}. ``LR'' stands for the learning rate. ``HQ'' stands for high-quality sources, as defined in Equation~\eqref{eqn:HQ_standard}. When a given set of parameters are not updated in a given phase, the initial learning rate is marked ``$-$''. In phase 1, we first train the model parameters ($\mathcal{W}$, $b$, $\vec{R}$) on HQ data while holding the stellar parameters ($\Theta$, $\varpi$, $E$) fixed. This produces an initial, rough model. We then train both the model and individual stellar parameters in phase 2, again using only HQ data. In phase 3, we apply the model trained with HQ data to all stars in the training set to refine their stellar parameters. In phase 4, we use a self-cleaning procedure to remove unresolved binary systems and other outliers. In phase 5, we use the remaining stars to train the model, while simultaneously updating the stellar parameters. At this point, the training of our model is complete. Finally, in phase 6, we use our trained model to update the parameters of all stars in the training set.}
    \label{tab:trainingproperties}
\end{table*}

Our goal is to find the set of both global model parameters and individual stellar parameters that maximize the above posterior density. In an ideal world, we would simultaneously infer all parameters -- both the global model parameters and the individual stellar parameters. However, we find an iterative approach to be more manageable. We implement our model in Tensorflow~2, and use gradient descent to iteratively update the global model parameters (holding the individual stellar parameters fixed) and the individual stellar parameters (holding the global model parameters fixed). Not all terms in the posterior need to be calculated for each type of gradient descent step -- we only need to calculate the terms that depend on the set of parameters being updated. Thus, when updating the global model parameters, we do not calculate the stellar priors or the likelihoods of $\hat{\varpi}$, $\hat{E}$ and $\hat{\Theta}$.

To train the global model parameters $\mathcal{W}$, $b$ and $\vec{R}$, holding the stellar parameters $\Theta$, $\varpi$ and $E$ fixed, we seek to maximize
\begin{align}
    p \left(
      \vec{f}_{\mathrm{obs}}
      \mid
      \mathcal{W}, b, \vec{R}, \Theta, \varpi, E, C_f
    \right)
    p\left(\mathcal{W}\right) p\left(b\right) p\left(\vec{R}\right)
    \, .
\end{align}
This is equivalent to minimizing $\chi_f^2$ plus an L2-norm penalty on $\mathcal{W}$:
\begin{align}
	\mathcal{L}_{\mathrm{model}}
	\left( \mathcal{W}, b, \vec{R} \right)
	&=
	\chi_f^2
	+ \frac{1}{n_W} \lVert \mathcal{W} \rVert_2
	  \, ,
	\label{eqn:model-loss}
\end{align}
with respect to the global model parameters $\mathcal{W}$, $b$, and $\vec{R}$. Here, $\chi_f^2$ is the sum of the flux likelihood for a single random star, calculated according to Equation~\ref{eqn:chi2}, and $n_W$ is the total number of weights in $\mathcal{W}$. The L2-norm penalty is equivalent to a Gaussian prior on the neural network weights, and causes our loss function to prefer smaller weights in $\mathcal{W}$, and thus simpler models. For unit weights, this prior would be the equivalent of adding one to $\chi_f^2$. In practice, during each gradient descent step, we calculate the average of $\mathcal{L}_{\mathrm{model}}$ for batches of 512 stars, in order to reduce the noise of the gradient.

To learn the parameters describing an individual star, holding the global model fixed, we seek to maximize all of the terms on the right-hand side of Equation~\eqref{eqn:model-posterior}, except for the last three prior terms, which do not depend on the individual stellar parameters. This is equivalent to minimizing the stellar loss function
\begin{align}
	\mathcal{L}_{\mathrm{star}}
	\left( \Theta, \ln\varpi, \ln E \right)
	&=
	  \chi_f^2
	  + \left( \frac{\varpi-\hat{\varpi}}{\sigma_{\varpi}} \right)^{\! 2}
	  \! + \left( \frac{E-\hat{E}}{\sigma_{E}} \right)^{\! 2}
	  \\ & + \sum_{i=0}^{3}
	      \left( \frac{\Theta_i-\hat{\Theta}_i}{\sigma_{\Theta,i}} \right)^{\! 2}
		-2\ln\varpi-2\ln E, 
	\label{eqn:stellar-loss}
\end{align}
with respect to the individual stellar parameters, $\Theta$, $\varpi$ and $E$. The 1st term corresponds to the likelihood of the observed fluxes (calculated using Equation~\ref{eqn:chi2}), while the 2nd, 3rd and 4th terms correspond to the likelihoods of $\varpi$, $E$ and $\Theta$. The last two terms correspond to the Jacobian terms $\frac{\partial \varpi}{\partial \ln\varpi}$ and $\frac{\partial \mathrm{E}}{\partial \ln\mathrm{E}}$, respectively.


We train our model and update the stellar parameters in stages. In the first stage, we fix $\Theta$, $\varpi$ and $E$ at their mean observed values, as determined by LAMOST, \Gaia and Bayestar19.

We use a stochastic gradient descent (SGD) optimizer with momentum of 0.5 to update the global model parameters. In order to update individual stellar parameters, we again use SGD, but with zero momentum. When alternating between different batches of stars, this prevents the gradient descent direction of one batch of stellar parameters from contaminating the gradient descent direction of the following batch of stellar parameters.

We train the model in several phases. In the first phase, we use a subset of stars with high-quality measurements to train our stellar flux and extinction model. In this phase, we update the global model parameters, while holding the stellar parameters fixed at their measured values (as determined by LAMOST, \Gaia and Bayestar19). We define the ``high-quality subset'' as the stars (in the training set) that pass all of the following cuts:
\begin{equation}
    \label{eqn:HQ_standard}
    \begin{aligned}\setcounter{mysubequations}{0}
        &\mysubnumber\quad \sigma_{\teff} < 200\,\mathrm{K}\\
        &\mysubnumber\quad \sigma_{\feh} < 0.2\,\mathrm{dex}\\
        &\mysubnumber\quad \sigma_{\logg} < 0.2\,\mathrm{dex}\\
        &\mysubnumber\quad \hat{\varpi} / \sigma_{\varpi} > 10\\
        &\mysubnumber\quad \sigma_E < 0.1
    \end{aligned}
\end{equation}
The high-quality subset contains 1,861,666 sources (90.1\% of the training set). Using this high-quality subset, we minimize $\mathcal{L}_{\mathrm{model}}$ by performing stochastic gradient descent on $\mathcal{W}$, $b$, and $\vec{R}$, or the ``red blocks'' in Fig.~\ref{fig:model-structure}. We use batches of 512 stars to calculate $\mathcal{L}_{\mathrm{model}}$ at each gradient descent step. In the following, we will use the concept of a training ``epoch,'' which is the number of training batches (or equivalently, training steps) required to run through the training dataset. For our training dataset, one epoch contains approximately 5,000 batches. In the first phase, we train for 128 epochs. We begin with a learning rate of $10^{-3}$, and reduce the learning rate by a factor of 10 at epochs 32, 64 and 96.

After this initial training phase, which produces a reasonable first guess of our stellar flux and extinction model, we proceed to a second training phase, in which we simultaneously refine both the global model parameters and the individual parameters of the stars in the high-quality subset. We again train for 128 epochs. In each training step, we first update the global model parameters, and then update the parameters of the current batch of stars. We begin with a learning rate of $10^{-5}$ for the global model parameters, and $10^{-4}$ for the individual stellar parameters, and reduce the learning rates by a factor of 10 at epochs 32, 64 and 96.

We would next like to train using the entire training set, including stars with larger measurement uncertainties. However, before using these stars to update the model, we would like to refine their individual parameters ($\Theta$, $\varpi$, $E$). We thus take 128 gradient descent steps for each star (including sources in the high-quality subset), holding the global model parameters fixed. We begin with a learning rate of $10^{-4}$, and again reduce the learning rates by a factor of 10 at epochs 32, 64 and 96.

Before proceeding to again update the global model parameters, we first carry out a self-cleaning step, in which we remove outlier stars based on large discrepancies in their measured vs. predicted fluxes or stellar parameters. We identify outliers in two ways. The first method is to remove any star for which our estimate of $\Theta$ or $\varpi$ is more than $4\sigma$ removed from the measured value. Binary systems, in particular, are likely a major contaminant in our dataset. A near-equal-mass binary will appear approximately twice as bright as would be predicted from the stellar type and parallax of the system. Our model will attempt to accommodate such systems by decreasing $\logg$ or increasing $\varpi$, both of which have the effect of increasing predicted flux. However, because LAMOST measures stellar atmospheric parameters using individual line shapes, rather than the overall luminosity of the source, it should not respond as drastically to the presence of a binary companion. We thus expect this self-cleaning method to identify (and remove) a large number of potential binary systems. This self-cleaning procedure is illustrated for low-extinction Solar-type stars in Fig.~\ref{fig:selfcleaning}. We define ``low-extinction Solar-type'' as:
\begin{equation}
\label{eqn:solar_type}
\begin{aligned}
    &|\teff-5700| < 50\ \mathrm{K},
    &\sigma_{\teff} < 100\ \mathrm{K}, \\
    &|\logg-4.5| < 0.1,
    &\sigma_{\logg} < 0.1, \\
    &|\feh-0.0| < 0.1,
    &\sigma_{\feh} < 0.1,\\  
    &|\sigma_{\varpi, \rm obs}/\varpi_{\rm obs}| < 0.1,\\
    &E < 0.05,
    &\sigma_E < 0.05
\end{aligned}
\end{equation}
 The second method of identifying outliers is to flag stars for which the residual between the observed and predicted flux at any wavelength is discrepant by more than $4\sigma$. Our parameter-based outlier rejection flags 21.4\% of sources, while our flux-based outlier rejection flags 6.9\% of sources. Because there is some overlap between these two populations, in total, we remove 24.7\% of sources through self-cleaning.

Next, we train both the model and refine the individual stellar parameters. We include all sources that pass our self-cleaning cut (1,556,666 sources, or 75.3\% of the training set) in this phase, and train for 128 epochs. In each training step, we first update the global model parameters, and then the parameters of the stars in the batch. We begin with a learning rate of $10^{-7}$ for the global model parameters, and $10^{-5}$ for the individual stellar parameters, and again reduce the learning rates by a factor of 10 at epochs 32, 64 and 96. This is the final phase in which we update the global model parameters.

Finally, we refine the individual stellar parameters, holding the global model parameters fixed. In this phase, use the Adam optimizer \citep{Kingma2014AdamOptimizer}, with an initial learning rate of 0.01, which we halve every 512 steps. We take a total of 4096 gradient descent steps for each star.

We summarize the key properties of each training phase in Table~\ref{tab:trainingproperties}. At the end of this process, we have a trained model of stellar flux and extinction as a function of wavelength, as well as an updated estimate of the type, parallax (or equivalently, distance) and extinction of each star.
    \section{Determination of stellar parameters}
\label{sec:stellar-parameter-determination}

We now wish to determine the parameters of all stars with XP spectra. Our approach is to use the model trained in the previous section to fit the stellar type, extinction and parallax, based on the observed flux and parallax. Although our optimization algorithm will be similar to the algorithm used to refine the stellar parameters during training in Section~\ref{sec:method}, 99\% of the BP/RP sources do not have observational constraints on stellar atmospheric parameters from LAMOST, and constraints on $E$ from Bayestar19 are unavailable for stars with ${\delta < -30^\circ}$. However, prior constraints on these parameters are still necessary, in order to prevent the optimizer entering regions of parameter space which are not physical, or which are not covered by the training data. We therefore replace the LAMOST constraints on stellar type by a Gaussian mixture model (GMM) prior. We additionally enforce positivity of extinction and distance by using $\ln E$ and $\ln \varpi$ as our parameters.

\subsection{Formulation and priors}
\label{sec:priors}

We determine the stellar parameters by maximizing the posterior probability of ($\Theta$, $\varpi$, $E$), given only the observed flux ($\vec{f}_\mathrm{obs}$, $C_f$) and observed parallax ($\hat{\varpi}$, $\sigma_{\varpi}$), under necessary constraints and priors. The global model parameters ($\mathcal{W}$, $b$, $\vec{R}$) have been fixed by the training process described in Section~\ref{sec:training}. If we omit constant factors, the posterior of a single star is given by
\begin{align}
  & p\left(
    \Theta, \varpi, E
    \mid
    \vec{f}_{\mathrm{obs}}, C_f, \hat{\varpi}, \sigma_{\varpi},
    \mathcal{W}, b, \vec{R}
  \right)
  \notag \\
  & \hspace{4em} \propto
  p\left(
    \vec{f}_{\mathrm{obs}}
    \mid
    \Theta, \varpi, E, C_f,
    \mathcal{W}, b, \vec{R}
  \right)
  \notag \\
  & \hspace{6em} \times
  p\left(
    \hat{\varpi} \mid \varpi, \sigma_{\varpi}
  \right)
  \notag \\
  & \hspace{6em} \times
  p\left(\varpi\right) p\left(E\right) p\left(\Theta\right)
  \, .
\end{align}

\begin{figure}
    \centering
    \includegraphics[width=1.0\linewidth]{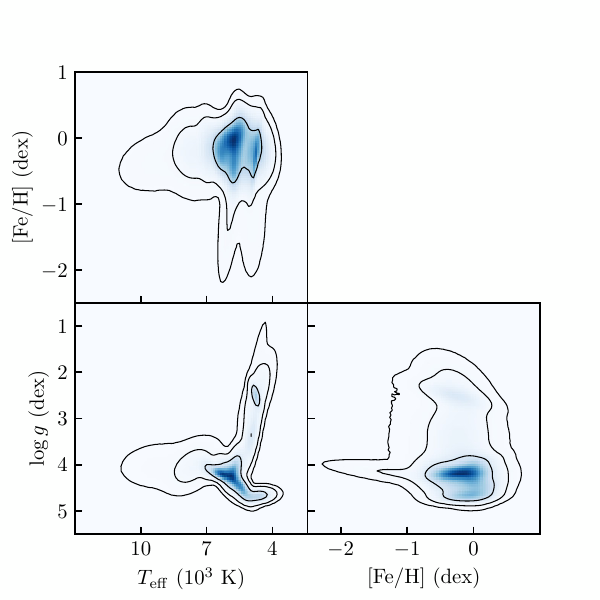}
    \caption{The prior assumption about the distribution of stellar types, represented by a Gaussian mixture model with N=16, with contours enclosing the 68\%, 95\%, 99\% percentiles. The Gaussian mixture model (GMM) is a linear combination of 16 independent normal distributions. As shown on the picture, the main sequence and the red clump stars are well represented, and regions are ruled out by the GMM, where the training set does not cover. This prior distribution prevents unrealistic solutions when optimizing stellar types, such as meaningless interpolations of parameters at locations where there are not enough stars in the training set.}
    \label{fig:stellarpriorsamples16components}
\end{figure}

The first term, $p\left(\vec{f}_{\mathrm{obs}} \mid \mathcal{W}, b, \vec{R}, \Theta, \varpi, E, C_f\right)$, is the likelihood of the observed flux, given by Equation~\ref{eqn:model-likelihood}. The second term is the likelihood of the observed parallax, which is a Gaussian with mean and standard deviation given by the GDR3 observation. We place a flat prior on positive parallaxes. As true parallax is strictly positive, we use $\ln \varpi$ as our variable (and include a corresponding Jacobian term in the prior). As \citet{bayestar19} does not cover $\delta<-30^\circ$, and as we want our determination of stellar extinction to be independent of \citet{bayestar19}, we impose a flat prior distribution of $\mathrm{E}$ on (0, $+\infty$). We use $\ln E$ as our variable, and include a corresponding Jacobian term ($-2\ln E$) in the prior. We use a Gaussian mixture model as the prior on atmospheric parameters ($\teff$, $\feh$, $\logg$), in order not to place them in regions of parameter space that are not covered by the training set (and for which we therefore do not have reliable models). The GMM is a probabilistic model describing an overall population (i.e. stellar type distribution in the training set) by the combination sub-populations (e.g., main sequence stars or the red clump), each with a Gaussian distribution in stellar-type-space:
\begin{equation}
    p_\mathrm{gmm}(\Theta)
    =
    \sum_{i=1}^{N}
    k_{i} \mathcal{N}\left(
      \Theta \mid \boldsymbol{\Theta}_{i}, \boldsymbol{C}_{i}
    \right)
    \, ,
    \label{eqn:gmm-prior}
\end{equation}
where $N$ is the number of sub-populations, and $\boldsymbol{\Theta}_{i}$, $\boldsymbol{\Sigma}_{i}$ and $k_i$ are the mean, covariance matrix and weight, respectively, of the $i^{\mathrm{th}}$ sub-population. The weights fulfill ${\sum_{i=1}^{N} k_{i} = 1}$. We use the \texttt{scikit-learn} package \citep{scikit-learn} to fit our GMM on the training set, and then implement the resulting model in TensorFlow~2, so that it is auto-differentiable. We use ${N=16}$ components to represent the stellar types in our training set. Our result is shown in Fig.~\ref{fig:stellarpriorsamples16components}. The main sequence and the red clump stars are well represented, and the regions sparsely covered by the training set are assigned low probabilities. By applying this GMM prior to the entire BP/RP spectral dataset, we implicitly assume that the stellar population of all BP/RP sources are well represented by the stars with LAMOST counterparts. Although this assumption is not true in detail, the purpose of our prior is primarily to prevent $\Theta$ from straying into regions of parameter space in which our model is unconstrained by data. In general, we expect the likelihood terms from the observed BP/RP spectra and parallax to dominate, and for the prior to play a subdominant role.

Combining all of our constraints, the loss function we finally use in the determination of stellar parameters is:
\begin{align}
    \mathcal{L}_{\text {inference}}(\Theta, \ln\varpi, \ln E)
    &=-2\ln(\text{Posterior}) \notag \\
    &=\chi_f^2 + \left(\frac{\varpi-\hat{\varpi}}{\sigma_{\varpi}}\right)^2
    -2\ln p_\mathrm{gmm}(\Theta) \notag \\
    &\hspace{1.2cm} -2\ln\varpi-2\ln E
    \label{eqn:opti_loss}
\end{align}
The first three terms correspond to the constraints from BP/RP flux, parallax and the prior of Gaussian mixture model, respectively. The last two terms are the Jacobian terms to convert variables from $\varpi$ and $E$ to $\ln \varpi$ and $\ln E$.

\subsection{Optimization of stellar parameters}
\label{sec:optimization}

Before optimizing the parameters of each star, we first conduct a rough grid search to locate a reasonable starting point. To construct our grid of possible starting points, we draw 512 samples from the Gaussian mixture model as the initial guess of stellar types, which is learned from the training set in section~\ref{sec:priors}. For each stellar type, we check a geometric series of extinction values, from $10^{-2}$ to $10$, with a common ratio of $10^{0.05}$. For each combination of stellar type and extinction, we calculate a parallax guess for each star by comparing the ratio of observed flux and predicted flux at 1~kpc. During calculation of this parallax guess, we only use fluxes between 592~nm and 782~nm, in order to avoid the edges of the BP/RP spectrum, which are typically noisy. Finally, for each star, we select the starting point $\left( \Theta, \varpi, E \right)$ that minimizes $\chi^2 - 2 p_{\mathrm{GMM}} \left( \Theta \right)$. 


After the starting points are determined, we use gradient descent to find the parameters that maximize the stellar posterior, which is equivalent to minimizing the loss function Equation~\eqref{eqn:opti_loss}. We separate our stars into batches of 32,768 ($10^{15}$), conducting several rounds of optimization of stellar parameters $\vartheta = (\teff, \feh, \logg, E, \varpi)$ on each batch. In the first round of optimization, we use the Adam optimizer, with an initial learning rate of 0.01 and 8,192 steps, halving the learning rate every 768 steps. For each round of optimization, we optimize the stellar parameters for 12288 steps to minimize the loss function of Equation~\ref{eqn:opti_loss}, in which we set the initial learning rate as 0.01, and reduce it by 8 every 512 steps. After each round of optimization, we compute the Hessian matrix of the loss for each star: 
\begin{align}
    H[{\mathcal{L}_{\text {inference}},\vartheta}]_{i,j}
    = \frac{
        \partial^2 \mathcal{L}_{\text {inference}}
      }{
        \partial \vartheta_i\partial \vartheta_j
      }
    \, .
\end{align}
We select the stars that have non-positive-semidefinite Hessian matrices for further optimization. In subsequent rounds of optimization, we use pure Stochastic Gradient Descent (SGD), with 12,288 steps per round. In the n$^{\mathrm{th}}$ round of SGD optimization, we use an initial learning rate of $10^{-5} / 2^\mathrm{n-1}$, halving the learning rate every 4,096 steps. If fewer than 0.01\% of the stars in a batch have non-positive-semidefinite Hessian matrices, we terminate optimization early and proceed to the next batch. Otherwise, we conduct at most 8 rounds of optimization for a given batch. Here we use (the eigenvalues of the) Hessian matrices as the criterion of convergence, because non-positive-semidefinite Hessian matrices can indicate that a minimum loss has not been reached.

\subsection{Estimation of uncertainties}
\label{sec:uncertainties}

\begin{figure}
    \centering
    \includegraphics[width=1.0\linewidth,trim=5 5 5 5,clip]{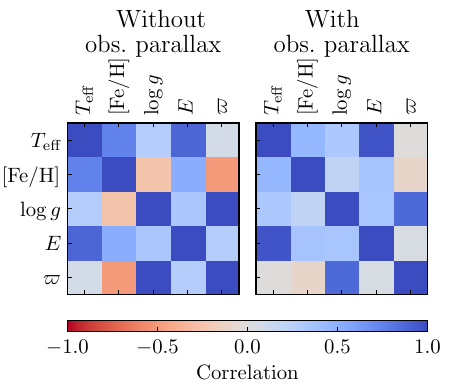}
    \caption{Observed parallaxes reduce the correlations among the stellar parameter estimates. The panels show the median correlations in the uncertainties of inferred stellar parameters before (left) and after (right) taking observed \Gaia parallaxes into account. The left panel only shows the correlations given by flux (\textit{i.e.} using only the first term in Eq.~\ref{eqn:fisher-theta}). The right panel contains the constraint from \Gaia's observed parallaxes (\textit{i.e.}, the second term in Eq.~\ref{eqn:fisher-theta}). Correlations between several pairs of parameters are significantly reduced by inclusion of a parallax observation: $(\varpi,\logg)$, $(\varpi,\feh)$, $(\varpi,E)$, $(\teff,\feh)$, and $(\feh,E)$. The strongest remaining correlation in the right panel is between $\teff$ and $E$, which both affect the slope of the spectra similarly.}
    \label{fig:param-correlations}
\end{figure}

We calculate uncertainties for our inferred stellar parameters using the shape of the likelihood function in the neighborhood of the best fit. In detail, we make use of the Fisher information matrix of the likelihood, which is related to the Hessian matrix, but which is guaranteed to be positive-semidefinite.

For our model, the Fisher information matrix of the combined flux and parallax likelihood is given by
\begin{align}
  \mathcal{I}_{\alpha,\beta} \left(\vartheta\right)
  &= \sum_{i,j}
    \frac{\partial f_{\mathrm{pred},i}(\vartheta)}{\partial \vartheta_\alpha}
    [C_f^{-1}]_{i,j}
    \frac{\partial f_{\mathrm{pred},j}(\vartheta)}{\partial \vartheta_\beta}
    \\ &\hspace{1cm}
    + \mathrm{diag} \left(0,0,0,0,\sigma_{\varpi}^2\right)_{\alpha,\beta}
  \, ,
  \label{eqn:fisher-theta}
\end{align}
where $\vartheta \equiv \left( \Theta, E, \varpi \right)$ is the full set of stellar parameters we fit, $C_f^{-1}$ is the inverse covariance matrix of the observed fluxes (see Eqs.~\ref{eqn:cov-obs}--\ref{eqn:cov-decomp}), and $\frac{f_{\mathrm{pred},i}(\vartheta)}{\partial \vartheta_\alpha}$ is the partial derivative of the $i^{\rm th}$ component of the predicted flux by the $\alpha^{\rm th}$ stellar parameter. As we implement our model in TensorFlow, these derivatives can be calculated quickly and accurately with automatic differentiation. The final term, which depends on $\sigma_{\varpi}^2$, represents the the additional information contributed by \Gaia's parallax measurement.

We evaluate the Fisher information matrix at the best-fit value of $\vartheta$ for each source. If our model $\vec{f}_{\rm pred}$ depended linearly on $\vartheta$, and if all of our measurements had perfectly Gaussian uncertainties, then the Fisher information matrix would be equal to the inverse covariance matrix of $\vartheta$. In general, the Fisher information matrix produces a \textit{lower bound} (the ``Cram\'{e}r-Rao bound'') on the parameter uncertainties. We use the Fisher information matrix as an estimate of our inverse covariance matrix, with the caveat that the true covariance matrix may encode somewhat larger uncertainties:
\begin{align}
    C_{\vartheta} \equiv
      \left[ \mathcal{I} \left(\vartheta\right) \right]^{-1}
    \, .
    \label{eqn:cov-theta}
\end{align}
Note that we do not take our GMM prior (see Eq.~\ref{eqn:gmm-prior}) on stellar type $\Theta$ into account when estimating our covariance matrix. This prior is intended mainly to prevent our inference from straying towards stellar types that are unphysical, or for which our model is unconstrained by training data. We therefore estimate our uncertainties solely on the basis of the flux and parallax likelihoods.

Fig.~\ref{fig:param-correlations} shows the median (over the entire XP catalog) Pearson correlation between each pair of stellar variables (where the correlation $\rho$ of variables $\alpha$ and $\beta$ is related to the covariance matrix $C$ by $\rho_{\alpha, \beta} = C_{\alpha, \beta}/\sqrt{C_{\alpha, \alpha}C_{\beta, \beta}}$), both before (left) and after (right) the constraints from \Gaia's observed parallaxes are incorporated into the uncertainty estimate (\textit{i.e.}, with and without the last term in Eq.~\ref{eqn:fisher-theta}).

If we rely on the XP spectra alone, there is a strong degeneracy between our inferred $\logg$ and $\varpi$. This is because the \Gaia XP spectral resolution is not fine enough to precisely measure linewidths. Holding other parameters constant, lower surface gravity corresponds to stars with larger radii, meaning that the major observable effect of $\logg$ on \Gaia XP spectra is to increase or decrease the overall luminosity of the star. Parallax, $\varpi$, also causes an overall scaling of apparent flux, leading to a strong degeneracy between $\varpi$ and $\logg$. Inclusion of the constraint from \Gaia's observed parallax lessens this degeneracy, decreasing the correlation coefficient between $\varpi$ and $\logg$ from 0.987 to 0.861. There are a number of other moderately strong degeneracies that are reduced by taking \Gaia's parallax measurements into account. Correlations are reduced between $\varpi$ and $\feh$ (from -0.497 to -0.098), $\varpi$ and $E$ (from 0.266 to 0.047), $\teff$ and $\feh$ (from 0.748 to 0.448), and $\feh$ and $E$ (from 0.515 to 0.352). The strongest remaining degeneracy in our model is between $\teff$ and $E$. This is because to first order, both parameters change the slope of the observed spectra. An increase in inferred temperature can be balanced by an increase in inferred reddening, rendering the color of the star nearly unchanged. This degeneracy is not lessened by inclusion of observed parallaxes.

\subsection{Validation of stellar parameters}
\label{sec:validation-stellar-parameters}

\begin{figure*}
    \centering
    \includegraphics[width=1.0\linewidth]{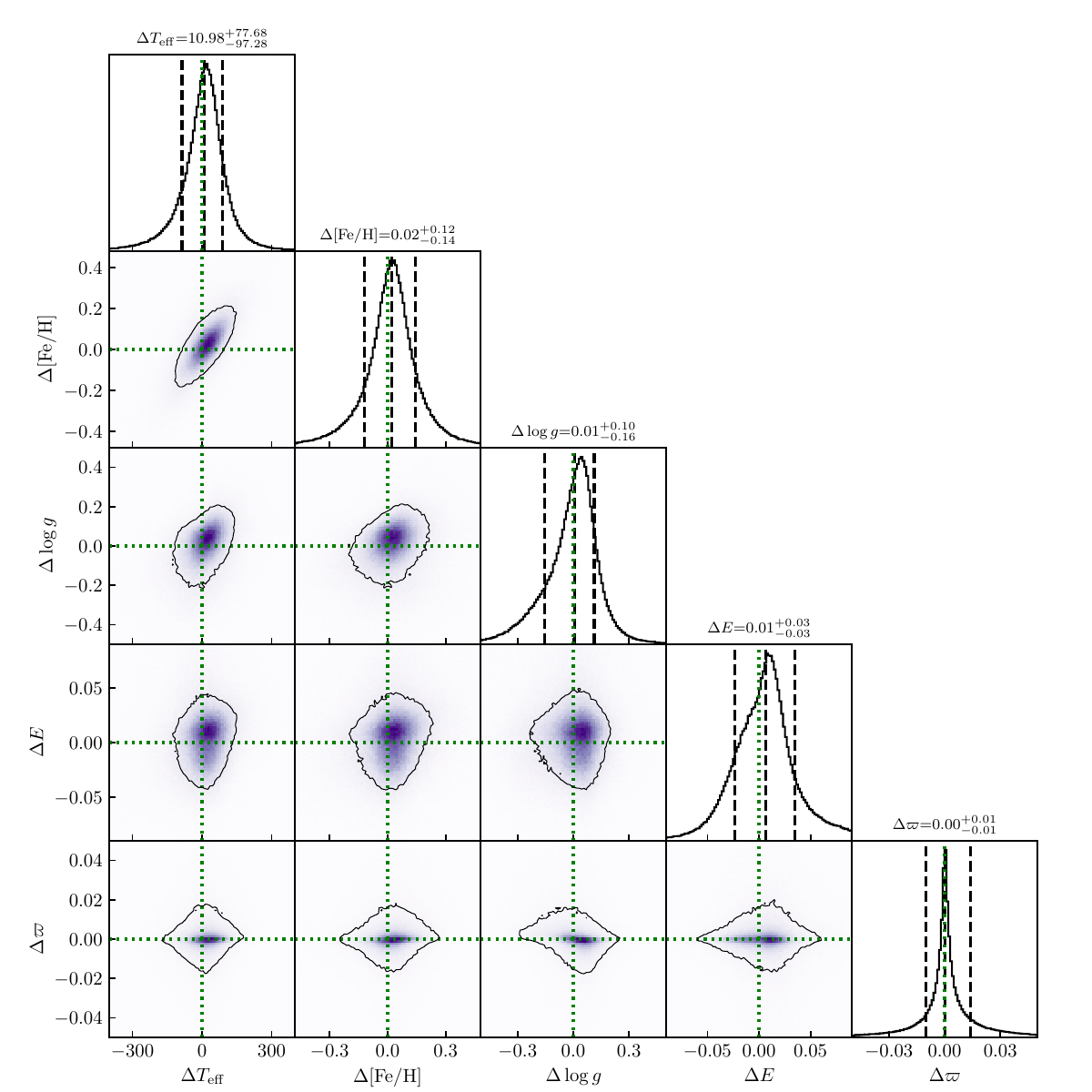}
    \caption{Difference between inferred stellar parameters $(\teff,\feh,\logg,E,\varpi)$ and the values determined by LAMOST, Bayestar19 and GDR3 in our validation dataset. We remove stars with bad parallax measurement or bad fitting results, which take $\sim3\%$ of all the stars. The contours enclose 68\% of the probability. These parameter estimates are determined using the loss function Eq.~\eqref{eqn:opti_loss}, which encodes a likelihood of the observed XP spectra and GDR3 parallax, with flat priors on $E$ and $\varpi$ and a simple Gaussian mixture model prior on $(\teff,\feh,\logg)$. Our model performs well on the validation dataset, obtaining nearly unbiased estimates of the parameters, with typical uncertainties of $\sim 90\,\mathrm{K}$ in $\teff$, $\sim 0.15\,\mathrm{dex}$ in $\feh$ and $\logg$, and $0.03\,\mathrm{mag}$ in $E$, although these uncertainties depend on the SNR of the observed XP spectra and parallaxes.}
    \label{fig:stellarparamresidualsrefineopt}
\end{figure*}

\begin{figure*}
    \centering
    \includegraphics[width=1.0\linewidth]{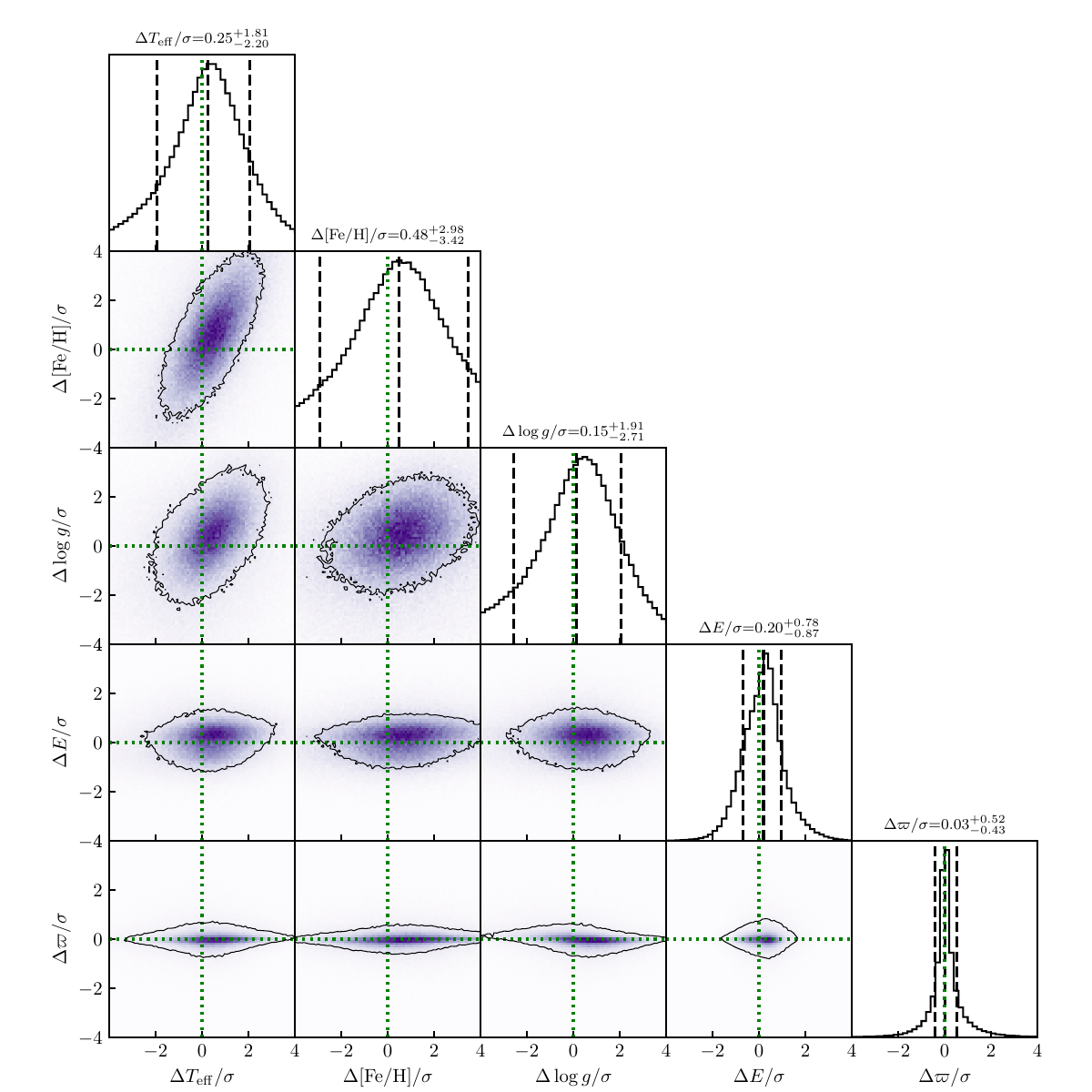}
    \caption{As Fig.~\ref{fig:stellarparamresidualsrefineopt}, but with residuals normalized by the uncertainties reported by LAMOST, Bayestar19 and GDR3. In our validation dataset, we find that our stellar parameters typically agree with those reported by LAMOST to within $1.5-3\sigma$. Specifically, We find that the deviations are of the same order of magnitude as those from LAMOST, Bayestar19 and GDR3:
        $\sigma_{\rm inference,\ \teff}\simeq2\sigma_{\rm LAMOST,\ \teff}$,
        $\sigma_{\rm inference,\ \feh}\simeq3\sigma_{\rm LAMOST,\ \feh}$,
        $\sigma_{\rm inference,\ \logg}\simeq2\sigma_{\rm LAMOST,\ \logg}$,
        $\sigma_{\rm inference,\ E}\simeq0.8\sigma_{\rm Bayestar19,\ E}$, and
        $\sigma_{\rm inference,\ \varpi}\simeq0.5\sigma_{\rm GDR3,\ \varpi}$.
    Our inferred uncertainties of $\varpi$ are even smaller than those reported by GDR3, because our parallax estimates are based on both the observed GDR3 parallaxes and the observed XP spectra.}
    \label{fig:stellarparamresidualsnormrefineopt}
\end{figure*}

\begin{figure*}
    \centering
    \includegraphics[width=1.0\linewidth]{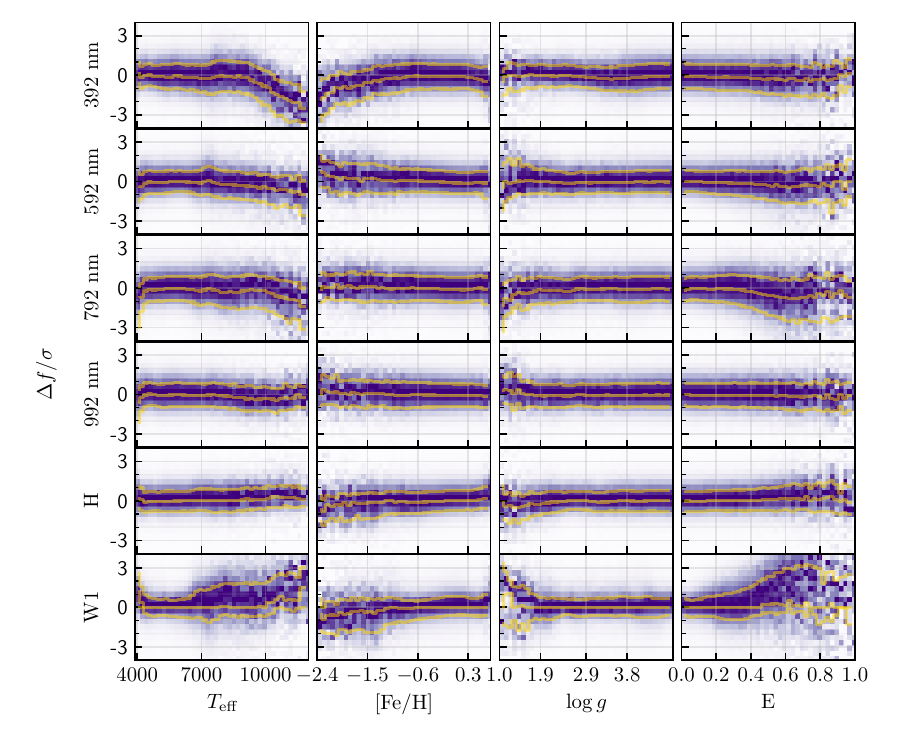}
    \caption{Normalized flux residuals ($\triangle f / \sigma \equiv \left( f_\mathrm{pred}-f_\mathrm{obs} \right) / \sigma_f$) in the validation set, as a function of $\teff$, $\feh$, $\logg$ and $\mathrm{E}$. The densities are normalized by the maximum value at each parameter value (\textit{i.e.}, in each pixel column). The yellow lines mark the position of the $15^{\mathrm{th}}$, $50^{\mathrm{th}}$ and $84^{\mathrm{th}}$ percentiles of the normalized residuals. During the optimization of parameters in the validation set, the model is unaware of the LAMOST and Bayestar19 estimates of stellar atmospheric parameters and extinction, respectively. The fluxes predicted by the model match the observed values to within $1\sim2\sigma$ over a wide range of parameter values, with the largest residuals occurring in $W1$ (particularly for $\teff \gtrsim 7000\,\mathrm{K}$ or $E \gtrsim 0.4$), and at 392~nm for $\teff \gtrsim 9000\,\mathrm{K}$.}
    \label{fig:residuals}
\end{figure*}

We validate our method by applying it to our validation set. We conduct the same optimization procedure discussed in Section~\ref{sec:optimization}, which does not take into account the LAMOST estimates of stellar type or the Bayestar19 reddening estimates. We then compare our inferred stellar parameters with those determined by LAMOST ($\Theta$), Bayestar19 ($E$), and GDR3 ($\varpi$). 
In order to estimate the typical errors with proper samples, we also remove outliers. Since the high-resolution spectroscopic constraints on $\log g$ are not available for most of the 220 million XP sources, we cannot apply the self-cleaning method used during training to remove suspected binary stars (see Section~\ref{sec:training} and Fig.~\ref{fig:selfcleaning}). We only remove stars with poorly constrained GDR3 parallax measurements ($\hat{\varpi}/\sigma_\varpi<5$) and the stars with bad fit quality ($\mathcal{L}_{\text{inference}}/{\text{DOF}}>5$), which altogether make up $\sim~3\%$ of stars in the validation set.

Fig.~\ref{fig:stellarparamresidualsrefineopt} shows the residuals between our inferred stellar parameters and the parameters reported by LAMOST, Bayestar19 and GDR3 (hereafter, the ``observed'' parameters), for stars in the validation set. The inferred and observed parameters match with typical errors of $\sigma_{\teff} \approx 90\,\mathrm{K}$, $\sigma_{\feh} \approx 0.15$, $\sigma_{\logg} \approx 0.15$, $\sigma_{E} \approx 0.03 \, \mathrm{mag}$, and $\sigma_{\varpi} \approx 0.01\,\mathrm{mas}$ (with $\sigma_{\varpi}$ depending strongly on distance).
We find that the errors are of the same order of magnitude as the observed uncertainties, as shown in Fig.~\ref{fig:stellarparamresidualsnormrefineopt}. We can therefore conclude that the parameters inferred using our model match the observations well, with no prominent biases, and that the typical residuals are comparable to the uncertainties of the observations.

We additionally estimate the quality of our predicted fluxes (evaluated at the parameter values inferred by our model) by comparison with the observed fluxes. Fig.~\ref{fig:residuals} shows the normalized flux residuals (prediction minus observation, divided by the uncertainty in the observation) in the validation set at 392, 592, 792 and 992~nm, and in 2MASS H- and unWISE W1-band, as a function of $\teff$, $\feh$, $\logg$ and $E$ (as estimated by LAMOST and Bayestar19). Over most of the parameter space, the model prediction matches the observation within $\pm 1 \sigma$.

\begin{figure}
    \centering
    \includegraphics[width=1.0\linewidth]{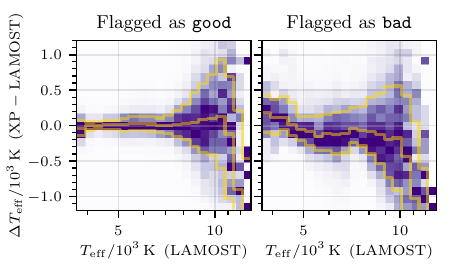}
    \includegraphics[width=1.0\linewidth]{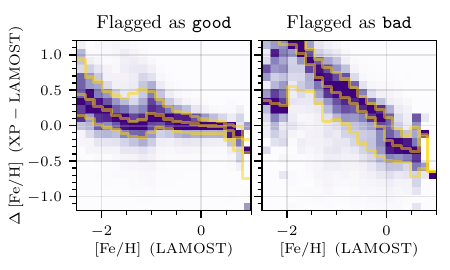}
    \includegraphics[width=1.0\linewidth]{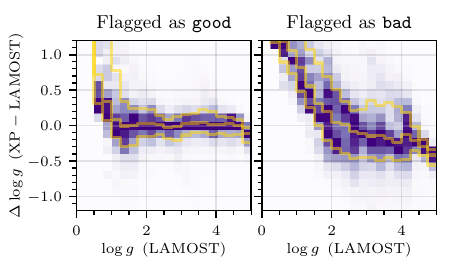}
    \caption{Residuals (model -- LAMOST) in $\teff$ (top panels), $\feh$ (middle panels) and $\logg$ for sources flagged as reliable (\texttt{good}, left panels) vs. unreliable (\texttt{bad}, right panels), as a function of the LAMOST atmospheric parameter estimates. The densities are normalized by the maximum value at each parameter value (\textit{i.e.}, in each pixel column). The yellow lines mark the positions of the $15^{\mathrm{th}}$, $50^{\mathrm{th}}$ and $84^{\mathrm{th}}$ percentiles of the residuals. The reliability classifiers are unaware of the LAMOST estimate, working instead with features that are available for all sources (including those without LAMOST observations). For each parameter, there is a stark difference between stars flagged as \texttt{good} and \texttt{bad}. There is a nearly flat trend in the median $\feh$ residuals of stars labeled \texttt{good} over a wide range of metallicities. $\logg$ estimates agree well for $1 \lesssim \logg \lesssim 5$, but agreement degrades for $\logg \lesssim 1$, where there is little LAMOST training data. $\teff$ residuals are approximately flat for $\teff \lesssim 11,000\,\mathrm{K}$, though the the scatter in the residuals increases dramatically for $\teff \gtrsim 7,500\,\mathrm{K}$.}
    \label{fig:validation-flags-lamost}
\end{figure}

\begin{figure}
    \centering
    \includegraphics[width=1.0\linewidth]{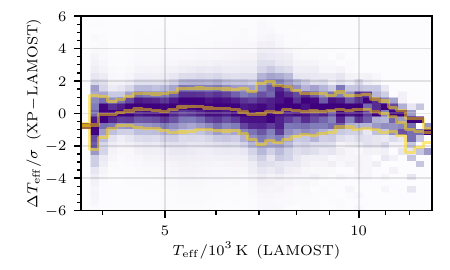}
    \caption{The distribution of normalized $\teff$ residuals (model minus LAMOST, divided by combined model and LAMOST uncertainties, as a function of LAMOST $\teff$. We include only estimates flagged as \texttt{good} by our $\teff$ reliability classifier. By comparison with the top-left panel of Fig.~\ref{fig:validation-flags-lamost}, we see that although the scatter in absolute $\teff$ residuals grows past $\teff \gtrsim 7,500\,\mathrm{K}$, the reported $\teff$ uncertainties also grow proportionally, leading to a well-behaved distribution of normalized residuals.}
    \label{fig:validation-teff-chi-lamost}
\end{figure}

Our stellar model does not cover all possible stellar types, and we therefore do not obtain reliable parameter estimates for all types of stars. In general, we recommend applying the following set of reliability cuts on our catalog of stellar parameters:
\begin{align}
    &\mathtt{chi2\_opt}/61 < 2
    \\
    &\mathtt{ln\_prior} > -7.43
    \\
    &\frac{\left| \mathtt{gaia\_parallax} - \mathtt{parallax\_est} \right|}{\mathtt{gaia\_parallax\_error}} < 10
    \label{eqn:reliable-fit}
\end{align}
For convenience of users of the catalog, we combine these cuts together into one flag, which we call \texttt{reliability}. The first cut corresponds to a cut on $\chi^2/\mathrm{dof}$, which captures whether the predicted spectrum matches the observed spectrum. The second cut ensures that the inferred stellar atmospheric parameters lie in a region that was covered by our LAMOST training dataset. The threshold on this cut is chosen so that 99.9\% of the probability mass of our prior on the stellar atmospheric parameters is enclosed in the allowed region. The third cut removes sources for which our parallax estimate strongly disagrees with that of \Gaia.

In addition to the above set of cuts, we develop a separate ``confidence'' measurement for each stellar atmospheric parameter, which provides the possibility of further ruling out spurious parameter estimates. We train a neural-network classifier for each atmospheric parameter ($\teff$, $\feh$, $\logg$), which predicts whether or not our estimate agrees well with LAMOST. We define agreement based on the metric
\begin{align}
    \chi_{\epsilon}^2 \equiv
    \frac{
      \left( \Delta x \right)^2
    }{
      \sigma_{\mathrm{opt}}^2 + \sigma_{\mathrm{LAMOST}}^2 + \epsilon^2
    } \, ,
\end{align}
where $\Delta x$ is the residual between our optimized parameter and LAMOST; $\sigma_{\mathrm{opt}}$ and $\sigma_{\mathrm{LAMOST}}$ are the uncertainties in our and LAMOST's parameter estimates, respectively; and $\epsilon$ is an uncertainty floor that we add in quadrature, so that sufficiently small residuals are considered to be in agreement. We use $\epsilon = 100\,\mathrm{K}$ for $\teff$, and 0.1~dex for $\feh$ and $\logg$. For each atmospheric parameter, we construct a training set by defining stars as \texttt{bad} when $\chi_{\epsilon}^2 > 9$, and as \texttt{good} when $\chi_{\epsilon}^2 < 4$. Our parameter estimates for this training set are derived without knowledge of LAMOST. During training, we withhold 20\% of the LAMOST sources to evaluate the performance of our flags. We only train this model on sources that pass our basic reliability cut (see Eq.~\ref{eqn:reliable-fit}). Our parameter estimates for sources which fail the basic reliability cut can be disregarded, regardless of our neural network ``confidence'' estimate.

As we would like to eventually apply these classifiers to the entire XP dataset, we can only take into account features that are available for every source (including sources not observed by LAMOST). These features include the reduced $\chi^2$ of our fit, the inferred $\teff$, $\feh$ and $\logg$, the \Gaia \texttt{parallax\_over\_error}, and the normalized flux residuals (model minus observed, divided by observational uncertainty) at every wavelength. More details about our classifier, including complete list of features used, are given in Appendix~\ref{sec:confidence-estimates}.

Each ``confidence'' classifier assigns a number between 0 and 1 to each source, which is the probability that our parameter estimate ($\teff$, $\feh$ or $\logg$) is reliable. We report these probabilities for every XP source. Fig.~\ref{fig:validation-flags-lamost} shows the performance of our confidence measurements on the 20\% of data withheld during training of the classifiers, considering only sources that pass our basic reliability cuts (Eq.~\ref{eqn:reliable-fit}). The distribution of $\teff$, $\feh$, and $\logg$ residuals (model -- LAMOST) is plotted on the $y$-axis for stars labeled \texttt{good} (left panels) vs. \texttt{bad} (right panels), using a threshold of 0.5 to distinguish the two classes. Along the $x$-axis, we separate stars into bins based their LAMOST parameter estimates. For all three atmospheric parameters, stars labeled \texttt{good} have much better behaved residuals than stars labeled \texttt{bad}. Fig.~\ref{fig:validation-teff-chi-lamost} focuses on the $\teff$ residuals, \textit{normalized} by the reported uncertainties. Although the absolute scale of the residuals are large for hot stars, the normalized residuals continue to be well behaved. Our model thus delivers highly uncertain $\teff$ estimates for hot stars, which are nevertheless statistically well behaved with respect to LAMOST. \change{In Table~\ref{tab:catalog-numbers}, we show the number of stars passing each quality flag.}

    \section{Results}
\label{sec:results}

This results section consists of three parts. First, we present our trained model, which maps stellar parameters ($\teff$, $\feh$, $\logg$, $E$, $\varpi$) to predicted \Gaia XP, 2MASS and WISE fluxes. Second, we apply our model to determine the stellar parameters of all 220 million XP sources in GDR3. Third, we discuss possible uses of our inferred stellar parameters. We show a preliminary three-dimensional dust map, based on our distance and extinction estimates, which shows evidence of unmodeled extinction curve variations. We also explore the metallicity distribution as a function of position in the Milky Way, based on our inferred stellar $\feh$ values.

\subsection{Trained model of stellar spectra and extinction}

\begin{figure*}
    \centering
    \includegraphics[width=1.0\linewidth]{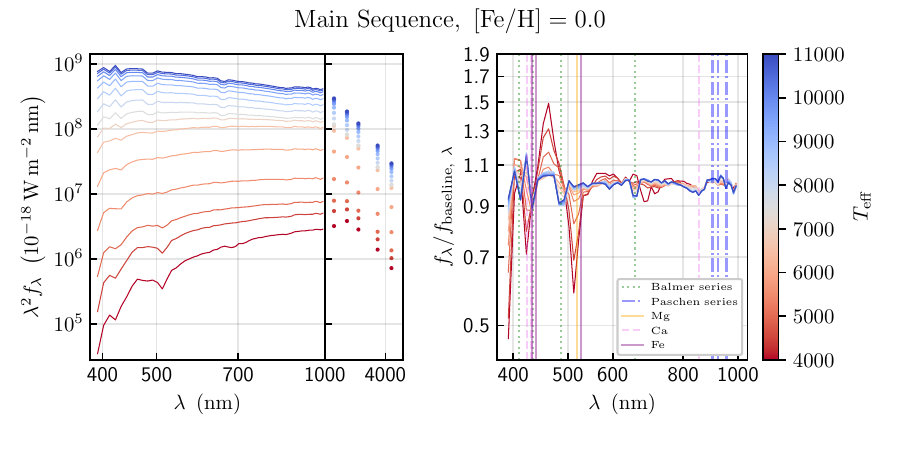}
    \caption{Model-predicted, zero-extinction fluxes (at 1~kpc) for the XP spectra and the NIR photometry (left panel), illustrated here for Solar-metallicity main sequence stars as a function of effective temperature. We set $\logg$ as a function of $\teff$ (see Eq.~\ref{eqn:logg-ms}). The right panel shows continuum-normalized XP model spectra to highlight the spectral line features. Balmer and Paschen series are shown as dotted and dash-dotted lines. We also show the position of Ca lines and Mg $\rm b_1$ line. As effective temperature increases, the flux increases, and Balmer lines become stronger. We also show the position of the Paschen series, and find that there is a weak trend of strengthening with the increase of $\teff$. It is clear that Ca lines around $\sim430$~nm and the Mg line attenuate as $\teff$ increase, but the trend for other metal lines are unclear, due to saturation and the low resolution of the XP spectra. Our full model, along with example code to evaluate it for arbitrary stellar types, extinctions and distances, is available at \url{https://doi.org/10.5281/zenodo.7692680}.}
    \label{fig:modelfluxvsdwarfs}
\end{figure*}

\begin{figure*}
    \centering
    \includegraphics[width=1.0\linewidth]{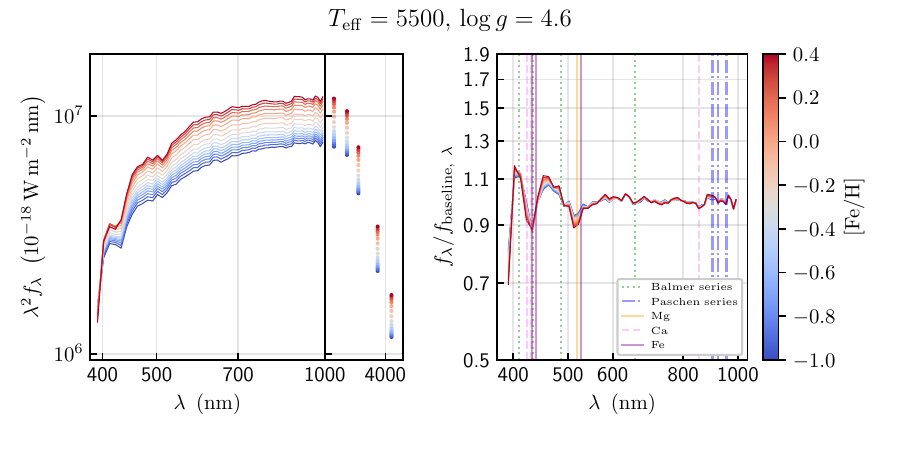}
    \caption{Predicted model fluxes as in Fig.~\ref{fig:modelfluxvsdwarfs}, here shown as a function of $\feh$ while fixing $T_{\text{eff}}=5500\,\mathrm{K}$ and $\log g=4.6$. As in Fig.~\ref{fig:modelfluxvsdwarfs}, the left panel shows spectral flux density for XP spectra and five NIR bands, while the right panel shows continuum-normalized XP spectra. As metallicity increases, Mg and Fe lines become deeper, while hydrogen lines do not vary significantly. There is also an overall increase of the flux with increasing $\feh$.}
    \label{fig:modelfluxvsfeh}
\end{figure*}

\begin{figure*}
    \centering
    \includegraphics[width=1.0\linewidth]{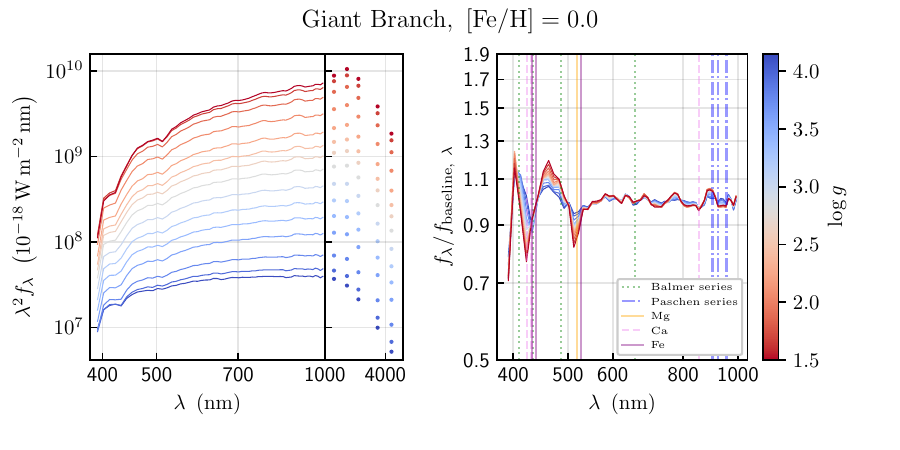}
    \caption{Predicted model fluxes as in Fig.~\ref{fig:modelfluxvsdwarfs}, here shown as a function of $\logg$ along the giant branch. For this figure, we use a piecewise function to set $\teff$ as a function of $\logg$ along the giant branch (see Eq.~\ref{eqn:teff-giants}). As in Fig.~\ref{fig:modelfluxvsdwarfs}, the left panel shows spectral flux density for XP spectra and five NIR bands, while the right panel shows continuum-normalized XP spectra. Because the absorption lines are not resolved, the primary effect of $\logg$ is to change the overall luminosity of the star. The variation of the continuum slope and line depths is due to the dependence of $\teff$ on $\logg$ along the giant branch.}
    \label{fig:modelfluxvsgiants}
\end{figure*}

We obtain a model that predicts \Gaia XP spectra (from 392-992~nm, with 10~nm resolution), as well as 2MASS and WISE ($W1$ and $W2$) bands, for any given stellar atmospheric parameters ($\teff$, $\feh$ and $\logg$), extinction ($E$) and parallax ($\varpi$). This requires our model to predict not only stellar absolute flux, but also to capture the wavelength-dependence of dust extinction.

Fig.~\ref{fig:modelfluxvsdwarfs} shows our model spectra for Solar-metallicity main-sequence stars from 4,000-11,000~K. In this figure, we set $\logg$ using a piecewise-linear function of $\teff$:
\begin{equation}
    \logg_{\rm MS}=\left\{
    \begin{array}{lc}
    4.6 \, , & {\teff < 5000 }\\
    4.6 \! - \! 5.0 \! \times \! 10^{-4} (\teff \! - \! 5000) \, , & {5000 \leq \teff < 6300}\\
    3.95 \, , & {\teff \ge 6300}\\
    \end{array} \right.
    \label{eqn:logg-ms}
\end{equation}
In the right panel of the figure, we plot the continuum-normalized model XP spectra, using a 5$^{\mathrm{th}}$-order polynomial to represent each spectrum's continuum.
Our model captures the temperature-dependence of both the slope of the flux continuum and of the depth of different classes of absorption lines. Although the low resolution of XP spectra makes it difficult to resolve individual absorption lines, a few classes of lines are nevertheless visible in our models. Both the Balmer and Paschen series appear in our models, with line strength increasing past $\teff \gtrsim 6000\, \mathrm{K}$, as expected, while features likely associated with Ca (at 423~nm) and Mg (at 523~nm) become stronger at lower temperatures. The Ca~II triplet (from 850-867~nm) is present as a single blended line across a wide range of temperatures.

Fig.~\ref{fig:modelfluxvsfeh} shows the dependence of our model spectra on $\feh$ for 5,500~K main-sequence stars. Increasing $\feh$ not only changes the overall scale of the flux, but also strengthens metal lines, as expected. As expected, the strength of hydrogen lines is insensitive to $\feh$. A feature likely associated with Mg is not cleanly centered at the expected location of the absorption line, at 523~nm, possibly because of the low resolution of our model (10~nm), and because this feature may be a combination of different lines.

Fig.~\ref{fig:modelfluxvsgiants} shows our model spectra as a function of $\logg$ along the giant branch. For plotting purposes, we define effective temperature along the giant branch as a piecewise-linear function of $\logg$:
\begin{equation}
    T_{\rm eff, GB}=\left\{
    \begin{array}{ll}
    5200 - 441.86 (3.65-\logg) \, , & {\logg < 3.65 }\\
    5900 - 1400 (4.15-\logg) \, , & {\logg \ge 3.65 }\\
    \end{array} \right.
    \label{eqn:teff-giants}
\end{equation}
As individual absorption lines are not fully resolved, the primary observable effect of $\logg$ in XP spectra is to change the overall luminosity of a star. The variation of the spectral shape in Fig.~\ref{fig:modelfluxvsgiants} is due to the relation between $\teff$ and $\logg$ along the giant branch.

\begin{figure}
    \centering
    \includegraphics[width=1.0\linewidth]{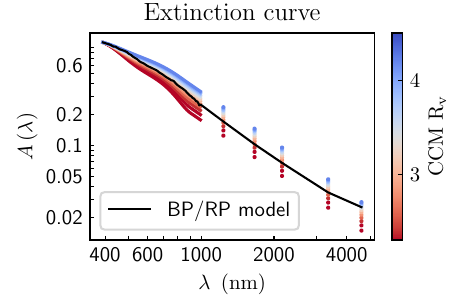}
    \caption{The extinction curve, assumed here to be universal across the sky, learned by our model (in black) from the data, compared with the CCM model \citep[colored curves;][]{ccm}. All models are normalized at $\lambda=392\rm\ nm$ Although we apply no constraints to ensure smoothness in the extinction curve, our model learns a curve that varies smoothly with wavelength. We find that our extinction model roughly aligns with CCM extinction curves with ${R_V \sim 2.8 - 3.2}$ in the XP spectral region, but deviates towards slightly larger values of $R_V$ in the NIR. Note, however, that the broad bandpasses of 2MASS and WISE cause extinction to vary by $\sim5\%$, compared to the mono-wavelength extinction. The full extinction curve is available as an electronic table at \url{https://doi.org/10.5281/zenodo.7692680}.}
    \label{fig:extcurvevsccm}
\end{figure}


Our model contains a universal extinction curve, which describes the relative extinction as a function of wavelength. Our model learns this extinction curve from the data alone, without reference to any previous extinction curve models. Each wavelength is modeled separately, without any prior on smoothness, meaning that the smoothness of the resulting extinction curve is a consequence purely of the data preferring such a curve. Fig.~\ref{fig:extcurvevsccm} shows our extinction curve (in black), and compares it with the family of $R_V$-dependent models of \citet[][hereafter ``CCM'']{ccm}. Our extinction model is smooth in the XP wavelength range, and roughly follows CCM models with ${R_V \sim 2.8 - 3.2}$. In the WISE W2 band, our extinction curve is slightly higher than the mean CCM curves, which corresponds to a deviation towards slightly higher values of $R_V$. This may be a reflection of the fact that our model predicts the extinction in the W2 band, which has a non-negligible spectral width. The extinction in the band is dependent on the slope of the stellar spectrum, which is generally sharply falling across W2 band. The band-weighted extinction is therefore expected to be larger than the monochromatic extinction at the center of the band.
able

\subsection{220 million stellar parameter estimates}
\label{sec:estimates_on_all_stars}

\begin{figure}
    \centering
    \includegraphics[width=1.0\linewidth]{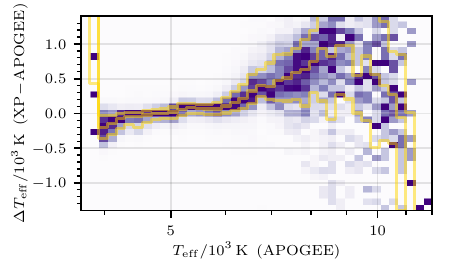}
    \includegraphics[width=1.0\linewidth]{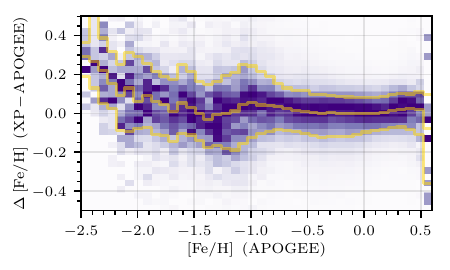}
    \includegraphics[width=1.0\linewidth]{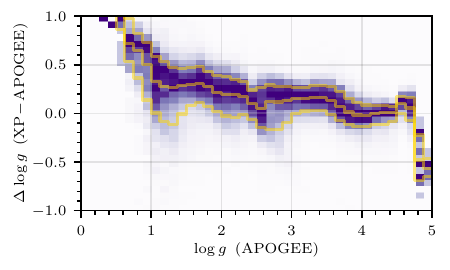}
    \caption{The distributions of residuals (model -- APOGEE) in $\teff$ (top panel), $\feh$ (middle panel), and $\logg$ (bottom panel), as a function of the respective APOGEE parameter estimates. The densities are normalized by the maximum value at each parameter value (\textit{i.e.}, in each pixel column). The yellow lines mark the positions of the $15^{\mathrm{th}}$, $50^{\mathrm{th}}$ and $84^{\mathrm{th}}$ percentiles of the residuals. In each panel, we plot only sources labeled as \texttt{good} by the respective reliability classifier for the given parameter (See Section~\ref{sec:validation-stellar-parameters}). We observe good agreement between our $\feh$ and $\logg$ estimates and those of APOGEE. However, we observe a trend in our $\teff$ residuals vs. APOGEE. This is due to two factors: a trend in LAMOST vs. APOGEE $\teff$ estimates, and large uncertainties in our $\teff$ estimates for hot stars.}
    \label{fig:validation-apogee}
\end{figure}

\begin{figure}
    \centering
    \includegraphics[width=0.96\linewidth]{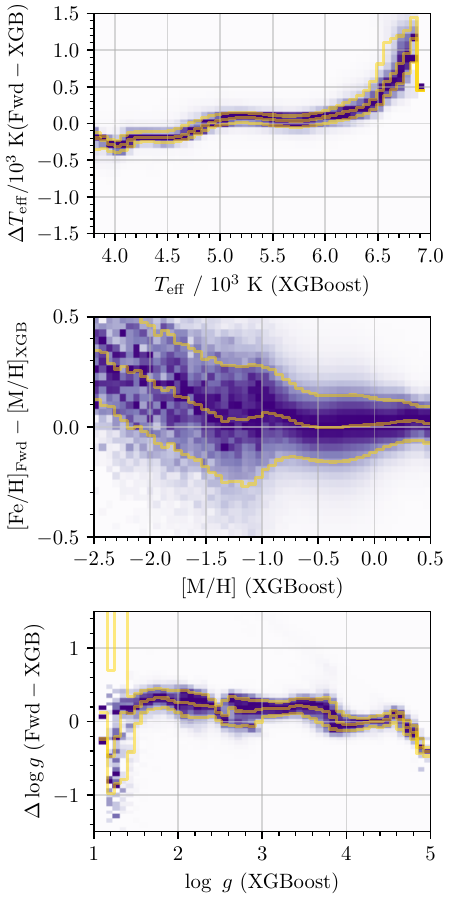}
    \caption{\change{The distributions of residuals (model -- XGBoost) in the same layout as Fig.~\ref{fig:validation-apogee}. ``Fwd'' represents the forward model we develop in this work, while ``XGB'' represents the XGBoost results from \citet{XGBoost_23}. We only compare the parameters of stars in which both catalogs have confidence. In our catalog, we select the stars passing the basic cut. In each panel, we additionally require that the corresponding paramter have $\mathrm{confidence} > 0.5$ (see Section~\ref{sec:validation-stellar-parameters}). In the XGBoost catalog, we select stars with parallax $\varpi>1\,\mathrm{mas}$, because the estimate of $\mathrm{[M/H]}$ in XGBoost is much less reliable for small parallax, according to Fig.~11 in \citet{XGBoost_23}. We find similar systematic trend in the $\teff$ residuals as we find in our comparison with APOGEE. This is due to the fact that XGBoost uses APOGEE as a training set, while we use LAMOST. A similar trend can be seen in a direct comparison of the underlying training datasets, LAMOST and APOGEE. Our $\feh$ and $\logg$ estimates have better agreement (with a scatter of approximately 0.2~dex), though we observe a systematic trend in the $\feh$ residuals for $\feh \lesssim -1.3$ and a systematic offset in $\logg$ of order 0.2~dex. As with the $\teff$ comparison, most of the systematic differences are due to disagreements between the underlying training datasets, LAMOST and APOGEE.}}
    \label{fig:validation-XGBoost}
\end{figure}

\begin{figure}
    \centering
    \includegraphics[width=1.0\linewidth]{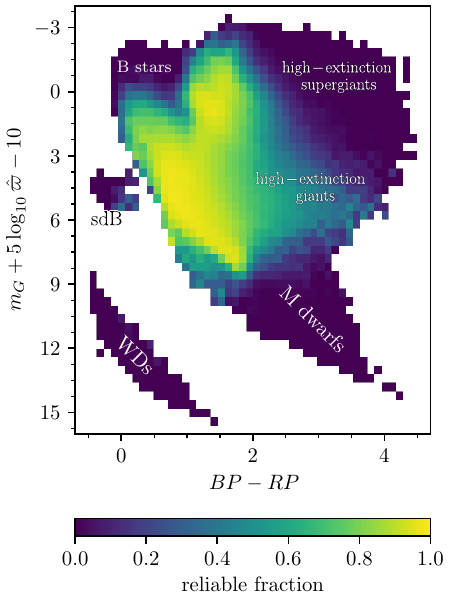}
    \caption{Fraction of sources that pass our standard reliability cut, as a function of position on the \Gaia CAMD. While we obtain reliable stellar parameters across the parameter range covered by our LAMOST training dataset, a number of classes of stellar objects lie outside this range: M-dwarfs, main-sequence B-stars, white dwarfs, B-type \change{subwarfs} (sdB), and supergiants. In addition, due to our assumption of a universal extinction curve, the reliability of our parameter estimates drops at large extinctions. This can be seen above for high-extinction giants. Use of additional sources of training data beyond LAMOST would expand the range of stellar types covered by our model.}
    \label{fig:reliable-frac-camd}
\end{figure}

We apply our model to the entire XP dataset, using the method described in Section~\ref{sec:stellar-parameter-determination}, to obtain stellar parameter estimates for 220 million sources, along with corresponding uncertainty estimates.

Our entire stellar parameter catalog is available for download at \url{https://doi.org/10.5281/zenodo.7692680}. We additionally plan to release the catalog through the Virtual Observatory in the near future. The columns of our catalog are described in Table~\ref{tab:catalog-columns}.

\begin{table*}
    \centering
    \begin{tabular}{c|c|p{0.6\linewidth}}
        Name & Data type & Description \\
        \hline
        \texttt{gdr3\_source\_id} & \texttt{integer} & \Gaia DR3 \texttt{source\_id}. \\
        \texttt{ra} & \texttt{float} & Right Ascension (in deg), as measured by \Gaia DR3. \\
        \texttt{dec} & \texttt{float} & Declination (in deg), as measured by \Gaia DR3. \\
        \texttt{stellar\_params\_est} & $5 \times \mathtt{float}$ & Estimates of stellar parameters ($\teff$ in kiloKelvin, $\feh$ in dex, $\logg$ in dex, $E$ in mag, $\varpi$ in mas). \\
        \texttt{stellar\_params\_err} & $5 \times \mathtt{float}$ & Uncertainties in the stellar parameters. \\
        \texttt{chi2\_opt} & \texttt{float} & $\chi^2$ of the best-fit solution. \\
        \texttt{ln\_prior} & \texttt{float} & Natural log of the GMM prior on stellar type, at the location of the optimal solution. \\
        \texttt{teff\_confidence} & \texttt{float} & A neural-network-based estimate of the confidence in the $\teff$ estimate, on a scale of 0 (no confidence) to 1 (high confidence). \\
        \texttt{feh\_confidence} & \texttt{float} & As \texttt{teff\_confidence}, but for $\feh$. \\
        \texttt{logg\_confidence} & \texttt{float} & As \texttt{teff\_confidence}, but for $\logg$. \\
        \texttt{quality\_flags} & 8-bit \texttt{uint} & The three least significant bits represent whether the confidence in $\teff$, $\feh$ and $\logg$ is less than 0.5, respectively. The $4^{\rm th}$ bit is set if $\mathtt{chi2\_opt}/61 > 2$. The $5^{\rm th}$ bit is set if $\mathtt{ln\_prior} < -7.43$. The $6^{\rm th}$ bit is set if our parallax estimate is more than $10\sigma$ from the GDR3 measurement (using reported parallax uncertainties from GDR3). The two most significant bits are always unset. We recommend a cut of $\mathtt{quality\_flags} < 8$ (the ``basic reliability cut''), although a stricter cut of $\mathtt{quality\_flags} == 0$ ensures higher reliability at the cost of lower completeness. \\
        \texttt{stellar\_params\_icov\_triu} & $15 \times \mathtt{float}$ & Upper triangle of the inverse covariance matrix of our stellar parameters. \\
        \texttt{stellar\_params\_cov\_triu} & $15 \times \mathtt{float}$ & Upper triangle of the covariance matrix of our stellar parameters, obtained from the inverse covariance matrix in a numerically stable manner that ensures positive semi-definiteness.
    \end{tabular}
    \caption{The columns of our stellar parameter catalog. The full catalog is available at \url{https://doi.org/10.5281/zenodo.7692680}.}
    \label{tab:catalog-columns}
\end{table*}

\begin{table*}
	\begin{tabular}{c|c|c}
		
		Quality cuts & \# of stars passing the cuts & Percentage \\ 
		\hline
		all stars & 219,197,643 & 100\%\\[1ex]
        
		Basic cuts & 180,344,401 & 82.3\%\\[1ex]

		Basic cut + Confident in $\teff$& 126,501,649 & 57.7\%\\[1ex]

		Basic cut + Confident in $\logg$& 138,890,532 & 63.4\%\\[1ex] 

		Basic cut + Confident in $\feh$& 132,573,594 & 60.5\% \\[1ex]  

		Basic cut + Confident in $(\teff,\feh,\logg)$& 90,390,241 & 41.2\% \\ 

	\end{tabular}
	\label{tab:catalog-numbers} 
	\caption{\change{Numbers of stars that pass different quality cuts. The ``basic cut'' is defined as $\mathrm{quality\_flags}<8$, as described in Section~\ref{sec:validation-stellar-parameters} and Table~\ref{tab:catalog-columns}. ``Confident in $\mathrm{param}$'' is defined as $\mathrm{confidence\_param}>0.5$.}}
\end{table*}


Table~\ref{tab:catalog-columns} describes the contents of our catalog. 82\% of our parameter estimates pass our basic reliability cut (see Eq.~\ref{eqn:reliable-fit}), which can be obtained from the catalog by requiring $\mathtt{quality\_flags} < 8$. Fig.~\ref{fig:reliable-frac-camd} shows the fraction of sources that are judged reliable by this cut, as a function of position on the CAMD. We obtain reliable parameter estimates for stars in the parameter range covered by our LAMOST training set. We do not obtain reliable parameter estimates for M-dwarfs, white dwarfs, B-type subdwarfs, or supergiants. Additionally, our reliability fraction drops at very large extinctions, likely due to variation in the extinction curve. Our model learns a universal extinction curve, and at very large extinctions, slight variations in the slope of the extinction curve can have a large effect on the ability of our model to ability to accurately model the stellar spectrum.

We additionally use the classifiers described in Section~\ref{sec:validation-stellar-parameters} to assign a ``confidence'' estimate to each estimate of $\teff$, $\feh$, and $\logg$. These classifiers assign greater than 0.5 confidence in our $\teff$, $\feh$ and $\logg$ estimates for 61\%, 68\% and 71\% of sources, respectively. A cut of $\mathtt{quality\_flags} == 0$ selects sources for which we are confident in $\teff$, $\feh$ and $\logg$, and which additionally pass our basic reliability cut (Eq.~\ref{eqn:reliable-fit}).

\subsection{External validation of stellar parameters}
\label{sec:outside-validation}

We validate our inferred stellar atmospheric parameters by comparison against an external catalog: APOGEE DR17. We crossmatch the XP sources with APOGEE DR17 stars, rejecting stars that have any of the following APOGEE flags set: \texttt{METALS\_BAD}, \texttt{SNR\_BAD}, \texttt{CHI2\_BAD}. When comparing each atmospheric parameter, we additionally require that our XP reliability probability be greater than 0.5 (See Section~\ref{sec:validation-stellar-parameters}). Fig.~\ref{fig:validation-apogee} shows the distributions of residuals in $\teff$, $\feh$ and $\logg$, as a function of the APOGEE measurement of each respective parameter. Our $\feh$ measurements agree to within 0.1--0.2~dex above $\feh \approx -2$, with somewhat larger scatter at lower metallicities. Our $\logg$ measurements agree to within a few dex for $\logg \gtrsim 1$, but with a slight trend for XP to overestimate $\logg$ relative to APOGEE. Temperatures agree well for $4,000\,\mathrm{K} \lesssim \teff \lesssim 6,000\,\mathrm{K}$, but show significant trends at higher temperatures. These trends are due to differences in LAMOST and APOGEE $\teff$ estimates. As shown in Figs.~\ref{fig:validation-flags-lamost} and \ref{fig:validation-teff-chi-lamost} our temperature estimates match those of LAMOST, though with large uncertainties for stars hotter than $\teff \approx 7,500\,\mathrm{K}$.

\change{We also compare our results with XGBoost \citep{XGBoost_23}, in which a discriminative model is applied to XP spectra to estimate stellar parameters, using APOGEE labels as training data. We only compare the parameters of stars in which both catalogs have confidence. In our catalog, we select the stars passing the basic cut and with $\mathrm{confidence} > 0.5$ in the parameter under consideration (see Section~\ref{sec:validation-stellar-parameters}). In the XGBoost catalog, we select stars with parallax $\varpi>1\,\mathrm{mas}$, because the estimation of $\mathrm{[M/H]}$ in XGBoost is less reliable for small parallaxes, according to Fig.~11 in  \citet{XGBoost_23}. Moreover, for stars with $\varpi<1\,\mathrm{mas}$ in the XGBoost catalog, we find a long-range distribution in $\logg$ for Red Clump stars, which we believe is because of the rapid degradation of discriminative models in low signal-to-noise regime. Fig.~\ref{fig:validation-XGBoost} shows our comparison with XGBoost. Our estimates of $\feh$ and $\logg$ generally agree well, with a typical scatter of 0.2~dex in most regimes. However, we find a systematic trend in the $\feh$ residuals for $\feh \lesssim -1.3$, and systematic offsets in the $\logg$ estimates of order 0.2~dex. We additionally find a sharp feature in the $\logg$ residuals near $\logg \approx 2.6$. We note that we find very similar features in our comparison with APOGEE, leading us to conclude that these systematic differences are primarily due to differences between the underlying training sets, LAMOST and APOGEE.}

\begin{figure*}
    \centering
    \includegraphics[width=1.0\linewidth]{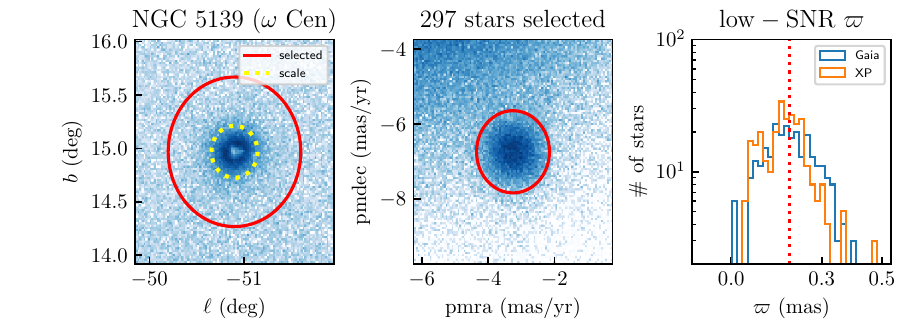}
    \includegraphics[width=1.0\linewidth]{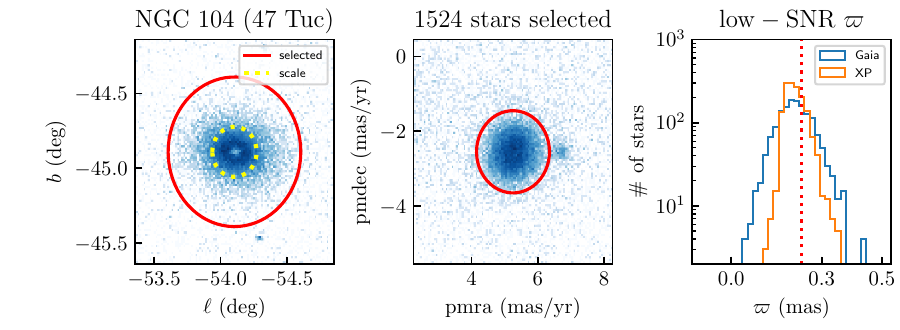}
    \includegraphics[width=1.0\linewidth]{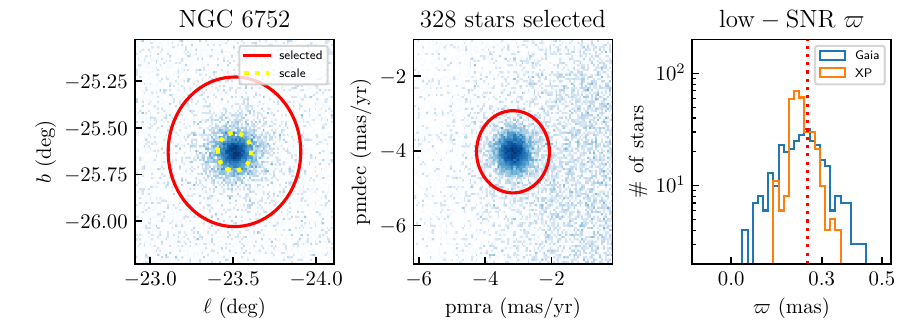}
    \caption{Validation of our parallax estimates in the low-SNR regime, using globular clusters of known distance as a test-bed. We use the three richest globular clusters, NGC~5139 (top row), NGC~104 (middle row) and NGC~6752 (bottom row) within 7~kpc. We identify stars in a globular cluster based on their positions in the sky (left panels) and in proper-motion space (middle panels). The solid red circles are the outer radius of our selection. In order to lessen the impact of crowding, which we expect to impact data quality, we exclude stars that are less than one ``scale separation'' (determined by \citealt{gc_catalog21}, and represented here by dotten yellow circles) from the center of the given cluster. We further require that stars have reliable parameter estimates (as defined by Eq.~\ref{eqn:reliable-fit}), and good confidence (> 0.5) on $(\teff, \feh, \logg)$, which are estimated in Section~\ref{sec:validation-stellar-parameters}.
    In the right panels, we show the distribution of the parallaxes, both as estimated by Gaia and as estimated with our model (using XP spectra). We conduct this comparison only for the low-SNR regime, defined as $\sigma(\varpi_{\rm obs}) > 0.2 \varpi_{\rm gc}$, where $\varpi_{\rm gc`}$ is the central parallax value of the given globular cluster, as reported by \citet{gc_catalog21}. In all three cases, our parallax estimates are more tightly concentrated about the central parallaxes (marked by red, dotted, vertical lines) of the respective globular clusters.}
    \label{fig:gc_parallax_comparison}
\end{figure*}

Our estimated parallaxes are constrained by both GDR3 parallax measurements and the observed stellar spectral energy densities, which are proportional to $1/\varpi^2_{\rm est}$. We therefore expect our estimated parallaxes to be closer to the true parallaxes, particularly in the regime in which GDR3 provides only weak constraints on parallax. Globular clusters provide a means of testing the validity of our parallax estimates, as many globular clusters lie at well known distances, and as relatively pure samples of cluster members can be straightforwardly obtained using simple cuts on sky location and proper motion. Our estimated parallaxes of the stars in a globular cluster should be more tightly concentrated around the central parallax, compared with the values observed by Gaia, particularly for stars with large GDR3 parallax uncertainties. At the same time, due to crowding, globular clusters are likely to have more contaminated XP spectra (particularly in the low-SNR regime that we are interested in), and thus present a relatively difficult test case for our method. In Fig.~\ref{fig:gc_parallax_comparison}, we show the three most populous globular clusters within 7~kpc: NGC 5139 ($\omega$~Cen), NGC~104 (47~Tuc) and NGC~6752 \citep{gc_catalog21}.
We restrict our comparison to stars with reliable parameter estimates, as identified by our basic reliability cut (Eq.~\ref{eqn:reliable-fit}) and good confidence (> 0.5) on ($\teff,\feh,\logg$), as described in Section~\ref{sec:validation-stellar-parameters}.
Because we wish to investigate the low-parallax-SNR regime, where XP spectra contribute proportionally more information about distance, we select stars with $\sigma(\varpi_{\rm obs}) > 0.2\varpi_{\rm gc}$, where $\varpi_{\rm gc}$ is the central parallax value of the given globular cluster, as reported by \citet{gc_catalog21}. As can be seen in the right panels of Fig.~\ref{fig:gc_parallax_comparison}, our estimated parallaxes are slightly more concentrated around the central values in all three globular clusters.

\begin{figure}
    \centering
    \includegraphics[width=1.0\linewidth]{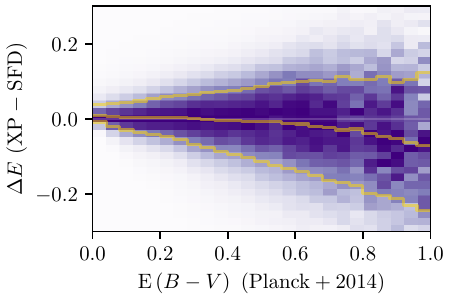}
    \caption{Comparison of our reddening estimates with SFD \citep{SFD1998}, for stars that lie outside of the Galactic midplane ($\left|z\right| > 400\,\mathrm{pc}$, $\left|b\right|>10\,\mathrm{deg}$) and outside the vicinity of the LMC and SMC, as a function of an independent measure of reddening, based on the Planck mission \citep{Planck2014dust}. The yellow envelopes mark the 16$^\mathrm{th}$, 50$^\mathrm{th}$, and 84$^\mathrm{th}$ percentiles of the reddening residuals. Our stellar reddening estimates agree well with SFD, with a median residual of approximately zero out to $\ebv \approx 0.5$, and a scatter of 15-20\%.}
    \label{fig:sfd-comparison}
\end{figure}

\begin{figure}
	\centering
	\includegraphics[width=1.0\linewidth, trim=0 10 0 51, clip]{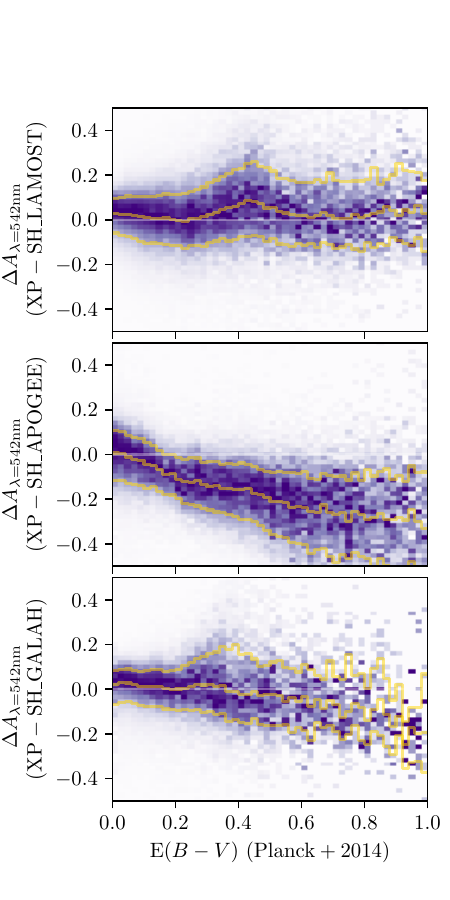}
	\caption{\change{Comparison of our reddening estimates with a recent \texttt{StarHorse} catalog \citep{Starhorse_23}. The layout is the same as in Fig~\ref{fig:sfd-comparison}. We compare the extinction estimates of stars in which both catalogs have confidence. In particular, we require that stars which pass the basic cut (see Section~\ref{sec:validation-stellar-parameters}) in our catalog, and that they have input flags of ``$\mathrm{panstarrs}$'', ``$\mathrm{2mass}$'' and ``$\mathrm{PARALLAX}$'' but no output flags in the \texttt{StarHorse} catalogs. \texttt{StarHorse} makes use of different spectroscopic catalogs. We find good consistency between our catalog and the \texttt{StarHorse} catalog using LAMOST LRS (top panel) and GALAH (bottom panel). However, there is a systematic trend when we compare our extinctions with those of \texttt{StarHorse} using APOGEE. We find a similar trend when we compares \texttt{StarHorse} estimates using LAMOST and APOGEE against one another. We believe that this is likely due to systematic differences in the $\teff$ estimates given by LAMOST LRS and APOGEE.}}
	\label{fig:sh-comparison}
\end{figure}

We validate our stellar reddening estimates by comparison with the \citet[][hereafter, ``SFD'']{SFD1998} dust map. Fig.~\ref{fig:sfd-comparison} shows the difference between our stellar reddening estimates (our parameter $E$) and the $\ebv$ estimate from SFD, as a function of reddening, as measured independently by the Planck mission, using dust optical depth at 353~GHz \citep{Planck2014dust}. As SFD only measures integrated reddening along the entire sightline, we restrict our comparison to stars which we expect to lie behind all of (or nearly all of) the dust. We therefore use stars that are more than 400~pc above or below the Galactic midplane (using our updated parallax estimate to calculate distance). We additionally exclude stars with $\left|b\right| < 10\,\mathrm{deg}$, which lie within 8~deg of the Large Magellanic Cloud or 6~deg of the Small Magellanic Cloud, or which fail our basic reliability cut (Eq.~\ref{eqn:reliable-fit}). Our stellar reddening estimates agree well with the reddening estimates provided by SFD, with only a slight trend in the median residuals past $\ebv \gtrsim 0.5$, and a scatter of 15-20\%.

\change{We also compare our extinction estimates with those given by the \texttt{StarHorse} catalogs \citep{Starhorse_23}. \texttt{StarHorse} infers stellar parameters from photometric, spectroscopic and astrometric data, using \textit{ab initio} stellar models. In  Fig~\ref{fig:sh-comparison}, we compare our extinction estimates (at $\lambda=542\,\mathrm{nm}$, denoted as ``$\mathrm{XP}$'') to those of \texttt{StarHorse}. \texttt{StarHorse} makes use of several different catalogs of spectroscopically determined atmosopheric parameters. We treat \texttt{StarHorse} estimates based on different spectroscopic surveys (LAMOST LRS, APOGEE and GALAH) separately (denoting the \texttt{StarHorse} estimates as \texttt{SH}\_\texttt{survey}). We find good agreement between our extinction estimates and those of \texttt{SH\_{LAMOST}} (top panel) and \texttt{SH\_GALAH} (bottom panel). However, there is a trend in the extinction residuals with \texttt{SH\_APOGEE}. We find a similar trend when comparing \texttt{SH\_APOGEE} with \texttt{SH\_{LAMOST-LRS}}. These differences may be related to differences between the temperature estimates produced by APOGEE and LAMOST LRS.}

\subsection{Preliminary map of extinction in 3D}

\begin{figure*}
    \centering
    \includegraphics[width=1.0\linewidth]{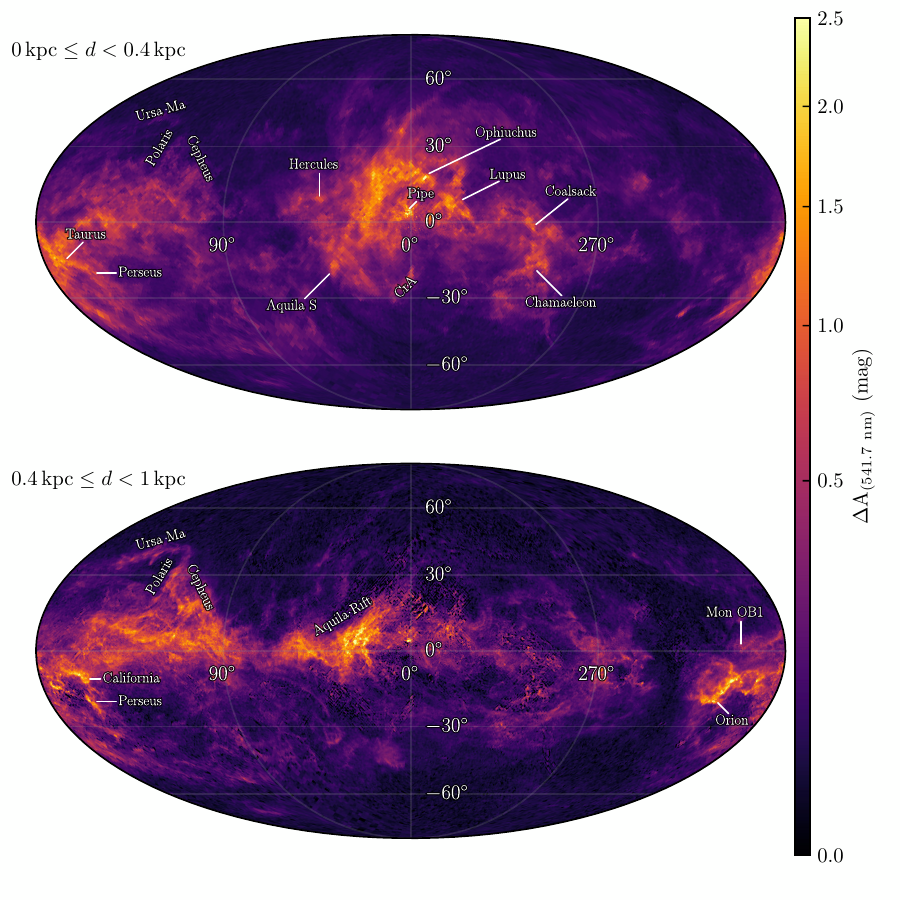}
    \caption{Sky maps of differential extinction in \change{different} distance ranges. We construct adaptive-resolution HEALPix maps of inverse-variance-weighted mean stellar extinction (using our inferred $E$ and $\sigma_E$) in each distance range, and subtract off the result for the previous distance range. We omit stars with poor fit quality ($\chi^2/\mathrm{DOF}>5$) or uncertain inferred parallaxes (defined as $\sigma_{\varpi}/\varpi>0.1$). We also require that $E<10$, $\sigma_E<0.04$, to remove outliers in extinction. This simple method of determining differential extinction is only intended to display the information present in our stellar reddening and distance inferences, and is not meant to replace (or reproduce) more sophisticated three-dimensional dust mapping methods. However, even this naive dust-mapping method recovers rich information on the three-dimensional structure of the interstellar medium, demonstrating the power of XP spectra to map dust. We recover the now-familiar structure of local dust clouds, several of which are labeled above.}
    \label{fig:dust-distance-slices}
\end{figure*}

\begin{figure*}
	\centering
	\includegraphics[width=1.0\linewidth]{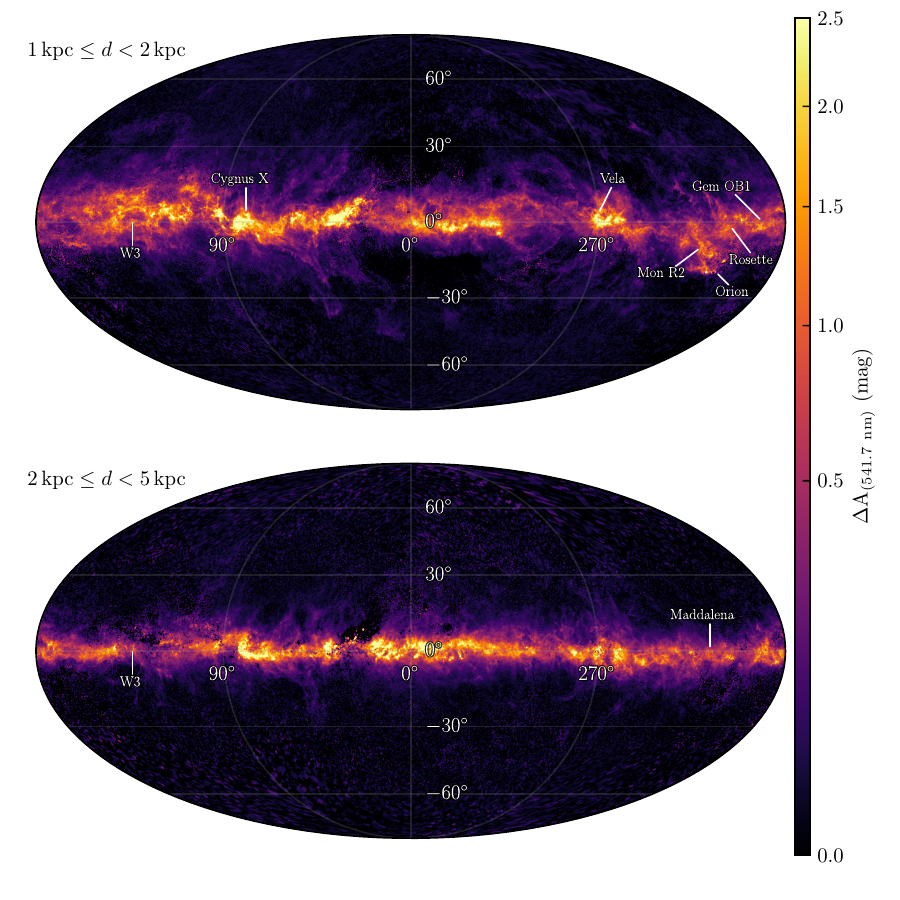}
	\caption{\change{Sky maps of differential extinction. As in Fig~\ref{fig:dust-distance-slices}, but for more distant slices.}}
	\label{fig:dust-distance-slices1}
\end{figure*}

\begin{figure*}
    \centering
    \includegraphics[width=1.0\linewidth]{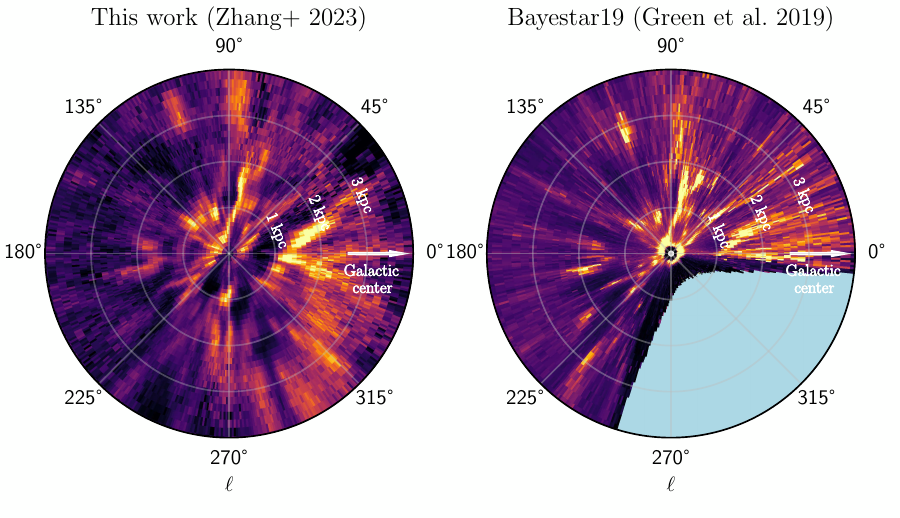}
    \caption{A bird's-eye view of the extinction density across the Galactic plane in our preliminary dust map (left) in Bayestar19 \citep[right;][]{bayestar19}. Both panels display average extinction density, integrated along the $z$-axis from $z = -400\,\mathrm{pc}$ to $+400\,\mathrm{pc}$). The Sun is located at ${(0, 0, 0)}$, with the Galactic Center off the plot to the right. Positions where Bayestar19 does not cover are marked as light blue. Using \Gaia XP sources alone, we are able to identify most of the dust structures in Bayestar19, which makes use of photometric extinction estimates for four times the number of stars. Because \Gaia surveys the entire sky, we also cover the Southern Galactic plane, which is not included in Bayestar19, due to the latter's reliance on Pan-STARRS~1, a photometric survey of the Northern Hemisphere. Even with extremely simple mapping techniques based on naive spatial binning, our XP-spectral-based extinction and distance estimates reveal a detailed three-dimensional dust distribution. More sophisticated mapping techniques hold the potential to extract much more detailed information from our stellar extinction and distance estimates.}
    \label{fig:dust-birds-eye}
\end{figure*}

Our final catalog of stellar parameters contains estimates of extinction ($E$) and parallax ($\varpi$) of all 220 million GDR3 XP sources, covering the entire sky. This is precisely the information required to produce a large-scale three-dimensional dust map. Compared to broad-band photometry, the XP spectra allow a much more precise determination of stellar extinction. While higher-resolution spectra would enable even more precise extinction estimates (by pinning down $\teff$, which is highly covariant with extinction), high-resolution spectral surveys do not presently observe enough stars to densely cover the sky and enable high-resolution, all-sky dust maps. With 220 million sources, the GDR3 XP catalog provides a unique combination of high sky density and precision extinction estimates. We therefore believe that \Gaia XP spectra will enable the next generation of three-dimensional Milky Way extinction maps. Here, we demonstrate a preliminary extinction map, based on naive spatial binning of stellar extinction estimates. We leave fuller exploitation of our stellar extinction estimates to future work.

\change{Figs.~\ref{fig:dust-distance-slices} and \ref{fig:dust-distance-slices1}} show preliminary sky maps of differential extinction in a number of distance ranges. In each distance bin, we calculate a HEALPix map of the inverse-variance-weighted mean extinction of stars falling in the bin (based on the inferred distance, $1/\varpi$, reddening $E$ and reddening uncertainty $1/\sigma_E$ of each star). In each distance bin, we calculate mean extinction maps at HEALPix resolutions of $\mathtt{nside} = 256$, 128 and 64 (equivalent to angular resolutions of $13.74^\prime$, $27.48^\prime$ and $54.97^\prime$, respectively), throwing out pixels that are based on fewer than 10 stars. We then combine these maps, using the highest-resolution map that covers any given part of the sky. After calculating a map of cumulative extinction in each distance bin, we calculate differential extinction by subtracting off the previous distance bin. It is this differential extinction that we display in Fig.~\ref{fig:dust-distance-slices}. We build this naive extinction map with the highest-quality stars: We remove stars with  $\mathcal{L}_{\text{inference}} / \mathrm{DOF}>5$. We also remove stars with extremely large extinctions ($E>10$), which are likely to be outliers, or with imprecisely determined extinctions ($\sigma_E>0.04$). Finally we remove stars with large distance uncertainties ($\sigma_{\varpi} / \varpi>0.1$, using our inferred parallax and its corresponding uncertainty). Although we do not impose spatial continuity on the dust, the rich structure of the ISM is already apparent, out to a distance of $\sim 5\,\mathrm{kpc}$. We mark the positions of several prominent dust clouds, including $\rho$~Ophiuchus, Aquila South, Hercules, Cepheus, Perseus, California, Ursa Major, Polaris, Taurus, Orion, the Pipe Nebula, Lupus, Chamaeleon, the Coalsack, Monoceros OB1, Cygnus X, Gemini OB1, Maddalena, W3, Rosette and Monoceros R2.

In Fig.~\ref{fig:dust-birds-eye}, we compare bird's-eye views of Galactic extinction density in our new catalog and in Bayestar19 \citep{bayestar19}. In detail, both panels of Fig.~\ref{fig:dust-birds-eye} show
\begin{align}
    A_{z=400\,\mathrm{pc}} \left( x, y \right)
    \equiv
    \frac{1}{800 \mathrm{pc}}
    \int^{400\ \mathrm{pc}}_{-400\ \mathrm{pc}}
        \frac{\mathrm{d}E(x,y,z)}{\mathrm{d}r}
        \, \mathrm{d}z,
    \label{eqn:birds-eye-extinction-integral}
\end{align}
where $r = (x^2+y^2+z^2)^{1/2}$ is distance from the origin in Sun-centered Galactic Cartesian coordinates, and $\mathrm{d}E / \mathrm{d}r$ is the extinction density (with units of $\mathrm{mag \, pc}^{-1}$). In order to calculate this integral, we need to construct an approximation of $\mathrm{d}E/\mathrm{d}r$ from our stellar distance and extinction estimates. To do so, we construct multi-resolution HEALPix maps of differential extinction in the same manner as described above for our sky maps of extinction, but with much finer distance bins.
Since this is only a naive average of stellar extinction ($E$), the precision of the naive dust map cannot exceed the uncertainties of distance measurements in GDR3, the lower limit of which increase linearly with distance. We use 40 bins to cover 0-5~kpc. The length of the n$^{\rm th}$ bin is $\sinh[n\cdot \text{arcsinh}(5)/40]-\sinh[(n-1)\text{arcsinh}(5)/40]$. We remove outlier stars with large uncertainties on $E$ ($\sigma_E > 0.04$), and stars with large uncertainties on parallax ($\sigma_{\varpi}/\varpi>0.33$). We further require that the stars have $\mathcal{L}_{\mathrm{inference}} / \mathrm{DOF} < 5$, to remove poorly fitting results.
We thus obtain a rough three-dimensional extinction map, $E(\ell,b,r)$. The differential extinction is then approximated by taking the difference in cumulative extinction between successive distance bins. We integrate the differential extinction through the Galactic plane vertically, using Eq.~\ref{eqn:birds-eye-extinction-integral}, to obtain our bird's-eye view of Galactic extinction.

While Bayestar19 makes use of a much larger stellar catalog than we use here, its stellar parameters are determined using broad-band photometry, rather than spectra. As can be seen in Figs.~\ref{fig:dust-distance-slices} and \ref{fig:dust-birds-eye}, using \Gaia XP sources alone, we are able to identify most of the prominent dust structures in Bayestar19, despite the difference in catalog size. Our stellar distance and extinction estimates build the foundation of a next-generation dust map.

\subsection{Flux residuals}
\label{sec:flux-residual-maps}

\begin{figure*}
    \centering
    \includegraphics[width=1.0\linewidth]{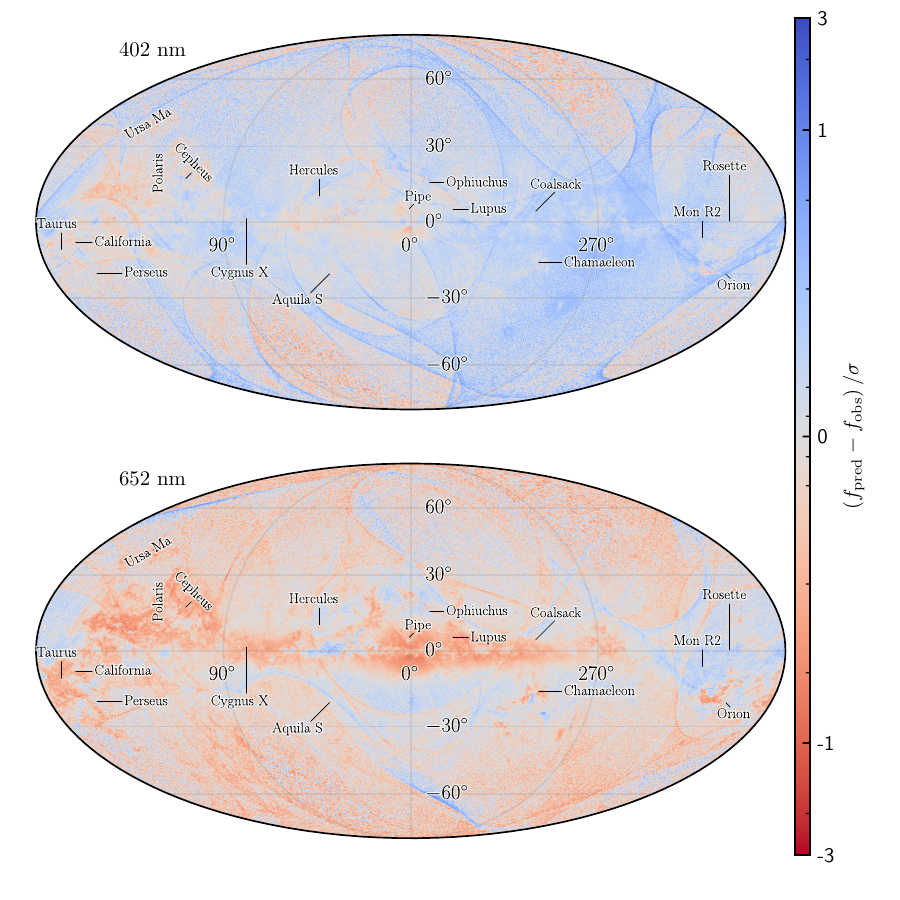}
    \caption{Maps of average model -- data flux residuals, normalized by observed flux uncertainties. In detail, we plot $\chi = (f_\mathrm{pred}-f_\mathrm{obs})/\sigma(f_\mathrm{obs})$ at 402~nm and 652~nm, and in 2MASS $J$ and WISE $W1$ band (see Fig~\ref{fig:valmapflux1}). We exclude the stars with \change{bad parallax agreement ($|\varpi_\mathrm{pred}-\varpi_\mathrm{obs}|/\sigma_{\varpi}>10$)}, poor fits ($\chi^2/\mathrm{DOF}>5$), \change{or which are not well represented in the training set ($\mathtt{ln\_prior} < -7.43$)}. In each panel, we also remove stars with extreme residuals ($|\chi|>10$) -- which are rare -- in the wavelength or band in question. With the exception of the WISE bands, average flux residuals are far smaller than typical flux uncertainties. However, several spatial patterns are visible in the residuals. \Gaia scanning-law patterns, which appear as arcs across the sky, can be seen at high Galactic latitudes in the 402 and 652~nm residual maps. \change{There are also faint dust-related residuals in the 652~nm map, visible both in the inner Galactic plane and in the vicinity of several well-known clouds (such as Orion, Cepheus, California and Taurus).}}
    \label{fig:valmapflux}
\end{figure*}

\begin{figure*}
    \centering
    \includegraphics[width=1.0\linewidth]{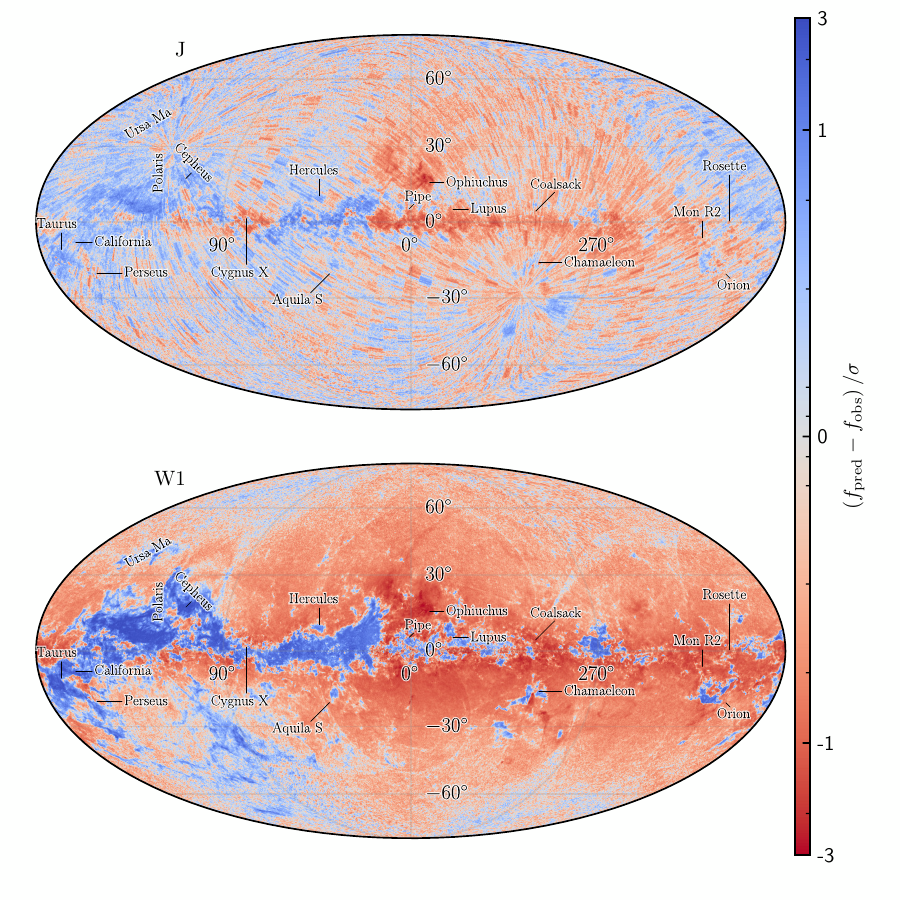}
    \caption{\change{Maps of average model -- data flux residuals, normalized by observed flux uncertainties. As in Fig~\ref{fig:valmapflux}, but for 2MASS $J$ and unWISE $W1$ bands. 2MASS exposure tiling patterns, which are aligned with equatorial coordinates, are visible in the $J$-band residuals. There are clear dust-related residuals in $W1$. These dust-related patterns tend to have opposite signs at short vs. long wavelengths (\textit{e.g.}, compare $W1$ here with 652~nm in Fig.~\ref{fig:valmapflux}), and have different signs in different clouds, suggesting that they are caused by variations in the extinction curve.}}
    \label{fig:valmapflux1}
\end{figure*}

Fig.~\ref{fig:valmapflux} and Fig.~\ref{fig:valmapflux1}  maps average flux residuals at 402~nm and 652~nm, and in 2MASS $J$ and unWISE $W1$ bands, normalized by the uncertainty of observations: $\chi \equiv (f_\mathrm{pred}-f_\mathrm{obs})/\sigma(f_\mathrm{obs})$. \Gaia scanning-law patterns are faintly visible in the average XP spectral flux residuals, while the 2MASS observing pattern is visible in the $J$-band residual map. However, the most prominent residuals are in the WISE bands, and clearly trace dust density. These patterns have two features which strongly suggest that they arise from unmodeled variations in the dust extinction curve. First, while the residuals clearly follow dust density, different clouds have residuals of opposite sign. For example, our predicted $W1$ fluxes are higher than observed fluxes in the Aquila Rift, Cepheus, and the Perseus-Taurus-Auriga complex, but are lower than observed fluxes in the $\rho$ Ophiuchi cloud complex. Such sign differences would arise if these two sets of clouds were to diverge from the mean $R_V$ in different directions. Second, the same residual patterns are visible to a lesser extent at shorter wavelengths (\textit{i.e.}, at 652~nm), but with the opposite sign. Changing $R_V$ alters the ratio of extinction at short and long wavelengths, and thus should lead to flux residuals of opposite sign on different ends of the observed spectrum. Given these factors (flux residuals that are proportional to dust density, but with different signs in different clouds; opposite flux residuals at short and long wavelengths), we hypothesize that these patterns are caused by extinction curve variations.

While it may appear counter-intuitive that dust-related residuals would appear most strongly at the longest wavelengths, this effect can arise if most of the likelihood constraints are from observations at shorter wavelengths, as in our case, where 61 of 66 observed wavelengths are in the optical range. Our model can compensate for unmodeled variations in the extinction curve by adjusting both $E$ and $\varpi$ in a way that nearly matches the observed fluxes at optical wavelengths, at the cost of introducing dust-dependent flux residuals in the NIR. In order to correctly model the spectrum across the entire observed spectrum, it is necessary to incorporate extinction curve variations into our model. We leave this to future work.

    \section{Discussion}
\label{sec:discussion}

Our model makes a number of assumptions:
\begin{enumerate}
    \item A star's intrinsic spectrum is a function of just three parameters: $\teff$, $\feh$, and $\logg$. At the coarse resolution of XP spectra, other parameters, such as $\left[\alpha/\mathrm{Fe}\right]$, rotation, or detailed chemical abundances can be ignored.
    \item The extinction curve is universal. That is, the extinction of any given star $i$ is a scalar multiple, $E_i$, of a universal function of wavelength, $R\left(\lambda\right)$.
    \item All observed spectra are of single stars. Our model does not account for binary systems.
    \item All observed spectra are stars in the range of stellar types covered by our model. Our model roughly covers the range $4000 \lesssim \teff \lesssim 10000$, and does not include \change{subdwarfs} and white dwarfs.
\end{enumerate}
We will discuss each of these assumptions in turn.

First, we model intrinsic stellar spectra as a function of $\teff$, $\feh$, and $\logg$. However, in our formalism, it is simple to include additional intrinsic stellar parameters, as long as training data is available in which these parameters have been measured. Given its use in measuring star formation history and the history of metal enrichment in the interstellar medium, $\left[\alpha/\mathrm{Fe}\right]$ would be a logical parameter to include in our model. However, studies using synthetic XP spectra suggest that for most stellar types, it is not possible to constrain $\left[\alpha/\mathrm{Fe}\right]$ precisely enough to be of interest \citep{Witten2022BPRPInformationContent,Gavel2021alphaFeBPRPExtraTrees}. Nevertheless, inclusion of $\left[\alpha/\mathrm{Fe}\right]$ as a fourth intrinsic stellar parameter would be a relatively straightforward addition to our model, requiring only a trivial change to the structure of our model (i.e., the addition of one dimension to $\Theta$; see Fig.~\ref{fig:model-structure}).


Second, this work is based on the assumption that the extinction curve is ``universal'' for all stars. That is, the extinction of any given star is proportional to a single, universal function of wavelength: ${A\left(\lambda\right) \propto R\left(\lambda\right)}$. We represent this function as a common 66-dimensional vector ($\vec{R}$), with each component representing a different wavelength. However, the Milky Way extinction curve is known to have non-negligible variation (\textit{e.g.}, \citealt{fitz86,fitz88,fitz90,fitz05,fitz07,fitz09}). As shown in Fig.~\ref{fig:extcurvevsccm}, the ratio of extinction at 900 and 400~nm ($R(\lambda\simeq900\mathrm{nm}) / R(\lambda\simeq400\mathrm{nm})$) can vary by as much as a factor of $\sim \! 2$. As discussed in Section~\ref{sec:flux-residual-maps} (see Fig.~\ref{fig:valmapflux}), we see evidence of extinction-curve variation in sky maps of flux residuals.

In the optical-through-near-infrared wavelengths, variation in the extinction curve can be effectively parameterized by a scalar, $R_V \equiv A_V/E(B-V)$ \citealt{ccm}. In the ultraviolet, which is not covered by our dataset, additional degrees of freedom may be essential to parameterize variation in the strength of the 2175~\AA{} bump and the slope of the far-UV rise in extinction \citep{PeekSchiminovich2013UVExtinction}. Therefore, the most practical way forward may be to add an additional degree of freedom to our extinction model, which allows the direction of $\vec{R}$ to vary. We believe that the XP spectra, in combination with 2MASS and WISE photometry, contain enough information to ``learn'' the variation of the extinction curve with minimal priors about the exact form of the variation. One possible way of implementing extinction curve variation in our model would be to expand the ``universal'' $\vec{R}$ in Equation~\eqref{eqn:full_model} to $\vec{R}+\xi\vec{R}_1$, where $\vec{R}_1$ is a vector orthogonal to $\vec{R}$, and $\xi$ is a scalar that is determined separately for each star. We would then expect $\xi$ to be related to $R_V$. This method is similar to that of \cite{sch16}, which decomposes the reddening vector into a series of orthogonal basis vectors. \citet{sch16} finds two statistically significant components, one of which represents a ``mean'' reddening curve, and the second of which is related linearly to $R_V$ (for small excursions from $R_V = 3.3$).

Although the specific physics behind $R_V$ variation are still not fully understood \citep{draine03}, the slope of the extinction curve is thought to be related to the dust grain-size distribution, which determines whether Rayleigh or Mie scattering applies. Larger dust grains lead to flatter curves, corresponding to larger $R_V$. A precise determination of $R_V$ as a function of position in the Milky Way would allow a study of the dependence of $R_V$ on environmental conditions, and would be of importance for the study of the dust properties as well as the evolution of the Galaxy. $R_V$ (or in our proposed model extension above, $\xi$) should vary with the physical properties of the dust, and should therefore be spatially continuous. This suggests the utility of using continuous function of position, $R_v(r,l,b)$, to model extinction curve variation. One way to implement this in our model would be to represent $R_V$ (or, equivalently, $\xi$) as a basis-function expansion, with the expansion coefficients being learned during training.

A third assumption made by our model is that all sources are single stars. However, a significant fraction of stars reside in binary systems. For large luminosity ratios, the observed flux will be dominated by the brighter star in the pair, and our model should reasonably recover the parameters of the brighter star. However, in binary systems in which the luminosity ratio is near order unity, we expect our model to break down. In the case of an unresolved, equal-mass binary system, in which both stars formed from the same cloud and thus have the same age and metallicity, both stars will be of equal temperature. The system will thus appear as a single star of twice the luminosity of a single star of the same ${\left( \teff, \feh, \logg \right)}$. In attempting to fit this system as a single star, our model could decrease the inferred distance by a factor of $\sqrt{2}$ (or equivalently, increase inferred parallax by a factor of $\sqrt{2}$), obtaining a solution matching the observed spectrum. The model also has an additional parameter, $\logg$, which it can decrease in order to increase the predicted luminosity of the observed spectrum. During training, we address the issue of binary stars using self-cleaning, as described in Section~\ref{sec:training}. This self-cleaning method makes use of the residuals between our inferred values of $\logg$ and $\varpi$, and those determined by LAMOST and Gaia. However, independently measured stellar atmospheric parameters are unavailable for the entire XP catalog (and indeed, if they were, they would render the need for stellar parameters determined from XP spectra moot), so this self-cleaning method cannot be applied to our entire catalog of inferred stellar parameters. A possible way forward here is to explicitly model each XP source as a binary system. This would not require a full doubling of stellar parameters, as binary systems can be reasonably assumed to share the same $\feh$, and as stellar evolution places constraints on the region of $\left( \teff, \logg \right)$-space that the two stars can simultaneously inhabit. Comparison of the goodness-of-fit (e.g., reduced $\chi^2$) of single-star and binary solutions for each source would then allow identification of possible binary systems.

\change{Finally, while our method is formulated in a way that is} agnostic towards the origin of the stellar parameters used during training, our stellar model can only be as good as the higher-resolution spectroscopic data used to train it. In this work, we have used LAMOST spectra, because of LAMOST's good coverage of the Hertzsprung-Russell Diagram, and of the main sequence, in particular. However, there are two drawbacks to this choice. First, LAMOST does not cover the giant branch as fully as APOGEE. Second, as discussed in Section~\ref{sec:lamost}, we use a combination stellar parameters from the standard LAMOST catalog and from the ``Hot Payne'', due to the former's difficulty in modeling hot stars. This leads to a discontinuity in our training data at $\teff \approx 7000\,\mathrm{K}$.

\change{In our future work, we plan to combine more spectroscopic surveys for better coverage of stellar parameter space. Possible options may include LAMOST MRS, APOGEE and GALAH. However, methods must be developed to explain and reconcile the systematic difference between catalogs (such as the discontinuity seen at $\sim$7000~K in our training dataset), which are caused by their use of different reduction pipelines and spectroscopic instruments.} \change{Another particularly} promising dataset for future work is the SDSS-V Milky Way Mapper (MWM, \citealt{MWM}), which is gathering high-resolution ($R\sim22,000$) NIR and medium-resolution ($R\sim2,000$) optical spectra of over 6 million stars. MWM targets are primarily at low Galactic latitudes (\textit{i.e.}, in the Milky Way disk). Compared to LAMOST, MWM will contain a relatively large number of stars at high extinction, making it a powerful dataset for studying variations in the extinction curve.

    \section{Conclusions}
\label{sec:conclusions}

In this paper, we determine stellar parameters ($\teff$, $\feh$, $\logg$, $E$, $\varpi$) for all 220 million \Gaia XP sources.

These stellar parameter determinations are based on an empirical forward model of \Gaia XP spectra and infrared photometric bands $J$, $H$, $\mathrm{K_s}$ from 2MASS and $W1$, $W2$ from unWISE. This model maps stellar atmospheric parameters ($\teff$, $\feh$, $\logg$), extinction ($E$) and parallax ($\varpi$) to predicted XP spectra and infrared photometry. The model is trained using the subset of the XP sources which have higher-resolution spectroscopy from LAMOST, and which thus have well determined atmospheric parameters. We additionally use extinctions from Bayestar19 and parallaxes from GDR3. The matched XP--LAMOST data is separated into training (80\%) and validation (20\%) sets. We then simultaneously learn a stellar model and refine the parameters of the stars in the training set. Our model generally recovers the flux of the stars in the validation set within $3\sigma$ in the regions of the parameter space where we have good data coverage. We implement our stellar model in an auto-differentiable framework, TensorFlow~2, which allows us to efficiently infer parameters of observed stars and propagate observational uncertainties into uncertainties in stellar parameters.

We apply this model to all 220 million published GDR3 XP spectra, 99\% of which do not have corresponding LAMOST spectra. When inferring the stellar parameters of all 220 million stars, we impose a weak prior on the stellar atmospheric parameters, based on the distribution of parameters in our training data. This is intended to prevent the optimizer from reaching regions of parameter space not covered by the training set. As our model is auto-differentiable, we determine the stellar parameters using simple gradient descent methods, and estimate the corresponding uncertainties using Fisher information matrices.

Our entire catalog of stellar parameters, along with our trained stellar model \change{and dust extinction curve}, are available \change{for download} at \url{https://doi.org/10.5281/zenodo.7692680}, \change{and can also be queried using ADQL/TAP from the German Astrophysical Virtual Observatory (GAVO; for details, see {\small \url{https://dc.zah.uni-heidelberg.de/tableinfo/xpparams.main}})}. \change{In order to obtain reliable stellar parameter estimates, we strongly urge users to apply our ``basic reliability'' cut: $\mathtt{quality\_flags} < 8$. We include additional ``confidence'' flags for each stellar atmospheric parameter, which provide even stricter cuts on the quality of our parameter inferences.}

Alongside the stellar atmospheric parameters, we obtain a large catalog of precisely determined stellar distances and extinctions, which will be the foundation of a next-generation 3D dust map in the Milky Way. We additionally find evidence that the \Gaia XP spectra contain information on variation in the dust extinction curve. We therefore expect ``the next generation'' of dust maps to benefit now only from the increased precision allowed by \Gaia XP spectra, but also to include variation of the extinction curve. We leave this avenue of investigation to follow-up work.

Finally, the stellar parameters presented in this paper are a promising resource for studying stellar populations in the Milky Way. Based on our validation dataset, we achieve typical errors of 90~K in $\teff$, 0.15~dex in $\feh$ and $\logg$, and 0.03~mag in $E$ (a parameter that is roughly equivalent to $\ebv$). 

    \section{Acknowledgements}

We would like to acknowledge the helpful conversions that we have had with colleagues on various aspects of this work. Morgan Fouesneau (MPIA) and Ren\'{e} Andrae (MPIA) helped us to understand the systematics and error properties of Gaia BP/RP spectra. David W. Hogg (NYU, CCA) provided advice on how to use matrix decompositions to deal with non-positive-semidefinite covariance matrices and to calculate Gaussian likelihoods in a numerically stable fashion, and advised us to provide inverse covariance matrices (as opposed to covariance matrices alone) in our final data release. Some of these conversations occurred during a hike up the Himmelsleiter in Heidelberg, a direct ascent of over 1200 stair steps. Gordian Edenhofer (MPA) suggested the use of Fisher information matrices, as a more numerically stable alternative to Hessian matrices, to estimate uncertainties on inferred stellar parameters. Edward F. Schlafly (STScI) provided helpful suggestions on how to diagnose flux residuals in the unWISE passbands. Douglas Finkbeiner (Harvard/CfA) and Andrew Saydjari (Harvard/CfA) provided feedback on our method and results in various discussions, including during hikes on the K\"{o}nigstuhl in Heidelberg. \change{Markus Demleitner (Astronomisches Rechen-Institut) uploaded and documented our stellar parameter estimates on GAVO.}

This work has made use of data from the European Space Agency (ESA) mission {\it Gaia} (\url{https://www.cosmos.esa.int/gaia}), processed by the {\it Gaia} Data Processing and Analysis Consortium (DPAC, \url{https://www.cosmos.esa.int/web/gaia/dpac/consortium}). Funding for the DPAC has been provided by national institutions, in particular the institutions participating in the {\it Gaia} Multilateral Agreement.

This work has made use of the Python package GaiaXPy (\url{https://gaia-dpci.github.io/GaiaXPy-website/}), developed and maintained by members of the Gaia Data Processing and Analysis Consortium (DPAC) and in  particular, Coordination Unit 5 (CU5), and the Data Processing Centre located at the Institute of Astronomy, Cambridge, UK (DPCI).

Guoshoujing Telescope (the Large Sky Area Multi-Object Fiber Spectroscopic Telescope, ``LAMOST'') is a National Major Scientific Project built by the Chinese Academy of Sciences. Funding for the project has been provided by the National Development and Reform Commission. LAMOST is operated and managed by the National Astronomical Observatories, Chinese Academy of Sciences.

Stellar parameter inference for the 220 million XP sources was carried out on the ``Raven'' HPC system, at the Max Planck Computing and Data Facility.
    \section{Data availability}
All underlying data used in this work is in the public domain, as detailed in Section~\ref{sec:data}. Our results can be acquired at \url{https://doi.org/10.5281/zenodo.7692680}, which contains all contents in Table~\ref{tab:catalog-columns}.
    
    \bibliographystyle{mnras}
    \bibliography{references} 
    
    \appendix
    \section{Gaia Archive Queries}
\label{sec:gaia-archive-queries}

In order to crossmatch 2MASS to GDR3 sources with XP spectra, we run the following ADQL query on the Gaia Archive:
\begin{lstlisting}[
           language=SQL,
           showspaces=false,
           basicstyle=\ttfamily\footnotesize,
           numbers=left,
           numberstyle=\tiny,
           commentstyle=\color{gray}
        ]
SELECT
  g.random_index, g.source_id,
  g.ra, g.dec,

  tmass.designation,
  tmass.ra as tm_ra, tmass.dec as tm_dec,
  tmass.j_m, tmass.j_msigcom,
  tmass.h_m, tmass.h_msigcom,
  tmass.ks_m, tmass.ks_msigcom,
  tmass.ph_qual,

  xmatch.angular_distance
    as gaia_tmass_angular_distance

FROM gaiadr3.gaia_source AS g
JOIN gaiadr3.tmass_psc_xsc_best_neighbour
AS xmatch
  USING (source_id)
JOIN gaiadr3.tmass_psc_xsc_join AS xjoin
  USING (clean_tmass_psc_xsc_oid)
JOIN gaiadr1.tmass_original_valid AS tmass
  ON xjoin.original_psc_source_id
     = tmass.designation

WHERE
  (g.has_xp_continuous = 'true')
  AND (
       (tmass.ph_qual LIKE 'A__')
    OR (tmass.ph_qual LIKE '_A_')
    OR (tmass.ph_qual LIKE '__A')
  )
  AND (tmass.ext_key is NULL)
  AND (g.source_id BETWEEN sid0 AND sid1)

ORDER BY g.source_id
\end{lstlisting}
Above, \texttt{sid0} and \texttt{sid1} are integers used to divide our query into manageable chunks.

We use the following query to fetch information from the GDR3 \texttt{gaia\_source} catalog and from the astrometric fidelity catalog \citep{gedr3fidelity} on each source with an XP spectrum:
\begin{lstlisting}[
           language=SQL,
           showspaces=false,
           basicstyle=\ttfamily\footnotesize,
           numbers=left,
           numberstyle=\tiny,
           commentstyle=\color{gray}
        ]
SELECT
  g.random_index, g.source_id, g.ref_epoch,

  g.ra, g.ra_error, g.dec, g.dec_error,
  g.parallax, g.parallax_error,
  g.pmra, g.pmra_error, g.pmdec, g.pmdec_error

  g.ruwe, g.astrometric_excess_noise,

  spur.fidelity_v2, spur.norm_dg,

  g.phot_g_mean_mag,
  g.phot_g_mean_flux, g.phot_g_mean_flux_error,
  g.phot_bp_mean_mag,
  g.phot_bp_mean_flux, g.phot_rp_mean_flux_error,
  g.phot_rp_mean_mag,
  g.phot_rp_mean_flux, g.phot_bp_mean_flux_error,

  g.has_xp_continuous,
  g.phot_bp_n_obs, g.phot_rp_n_obs,
  g.phot_bp_rp_excess_factor,
  g.visibility_periods_used

FROM gaiadr3.gaia_source as g
  JOIN gaiadr3.astrophysical_parameters as p
    ON g.source_id = p.source_id
  LEFT OUTER JOIN external.gaiaedr3_spurious as spur
    ON g.source_id = spur.source_id

WHERE
  (g.has_xp_continuous = 'True')
  AND (g.source_id BETWEEN sid0 AND sid1)

ORDER BY g.source_id
\end{lstlisting}
    \section{Stellar atmospheric parameter confidence estimates}
\label{sec:confidence-estimates}

As described in Section~\ref{sec:validation-stellar-parameters}, we train a classifier for each stellar atmospheric parameter ($\teff$, $\feh$, $\logg$), which assigns a ``confidence'' between 0 and 1 to each parameter estimate. Here, we list the input features used by the classifiers, and describe the neural-network structure of the classifiers, both of which are uniform across all three classifiers.

We use the following features:
\begin{itemize}
    \item \texttt{ln\_rchi2\_opt}
    \item \texttt{ln\_dplx2}
    \item \texttt{ln\_prior}
    \item \texttt{teff\_est}
    \item \texttt{logg\_est}
    \item \texttt{feh\_est}
    \item \texttt{teff\_est\_err}
    \item \texttt{logg\_est\_err}
    \item \texttt{feh\_est\_err}
    \item \texttt{asinh\_plx\_snr}
    \item \texttt{asinh\_g\_snr}
    \item \texttt{asinh\_bp\_snr}
    \item \texttt{asinh\_rp\_snr}
    \item \texttt{ln\_phot\_bp\_rp\_excess\_factor}
    \item \texttt{phot\_g\_mean\_mag}
    \item \texttt{ln\_ruwe}
    \item \texttt{fidelity\_v2}
    \item \texttt{norm\_dg}
    \item \texttt{ln\_bp\_chi\_squared}
    \item \texttt{ln\_rp\_chi\_squared}
\end{itemize}
Features of the form \texttt{asinh\_X\_snr} are computed using $\mathrm{asinh}\left(X/\sigma_X\right)$, using the GDR3 parallax and flux measurements. Features of the form \texttt{ln\_X} are calculated as $\ln \big[ \max \big( X, 10^{-7} \big) \big]$. The \texttt{ln\_dplx2} feature is computed from $\left| \varpi_{\mathrm{GDR3}} - \varpi_{\rm est} \right| / \sigma_{\varpi,\mathrm{GDR3}}$. The fields \texttt{fidelity\_v2} and \texttt{norm\_dg} are indicators of the quality of the GDR3 astrometric measurements and of crowding, respectively \citep{gedr3fidelity}. Features of the form \texttt{X\_est} and \texttt{X\_est\_err} are based on our stellar parameter estimates and their associated uncertainties. In addition to the above features, we additionally provide the network with information about the flux residuals between our best-fit models and the observed XP spectra. In detail, we provide the network with $\mathrm{asinh} \big[ \big( f_{\rm pred} - f_{\rm obs} \big) / \sigma_f \big]$ at each sampled wavelength.

We concatenate all of our input features into a single vector for each source, and feed it into a feed-forward neural network with three densely-connected hidden layers, each with 64 neurons and ReLU activation, and a single output neuron with a sigmoid activation. In order to regularize the network during training, we use a dropout fraction of 0.1 before each hidden layer \citep{Srivastava2014Dropout}, and impose an L2 penalty of $10^{-3}$ on the weights of each hidden layer. The output neuron represents the probability of the given stellar atmospheric parameter estimate being reliable. We use the binary cross-entropy between our output neuron and the training label as our loss function. We train each classifier using the Adam optimizer \citep{Kingma2014AdamOptimizer}, with a batch size of 4096, and an initial learning rate of $10^{-3}$, which we decrease by a factor of 10 whenever the validation loss fails to decrease over a span of 256 training epochs. We terminate the training procedure when the validation loss fails to decrease over a span of 1024 epochs.
    
    \bsp	
    \label{lastpage}
\end{document}